\documentclass[pldi]{sigplanconf}
\usepackage{enumitem}
\usepackage{amsmath}
\usepackage{amsthm}
\usepackage{color}
\usepackage{calc}
\usepackage{xspace}
\usepackage{upgreek}
\usepackage{graphicx}
\usepackage{natbib}
\usepackage{alltt}
\usepackage{array}
\usepackage{tabu}
\usepackage{multirow}
\usepackage{mathpartir}
\usepackage{balance}
\usepackage{amssymb}
\usepackage{ifxetex}
\usepackage[firstpage]{draftwatermark}
\SetWatermarkText{\hspace*{8in}\raisebox{6in}{\includegraphics[scale=0.1]{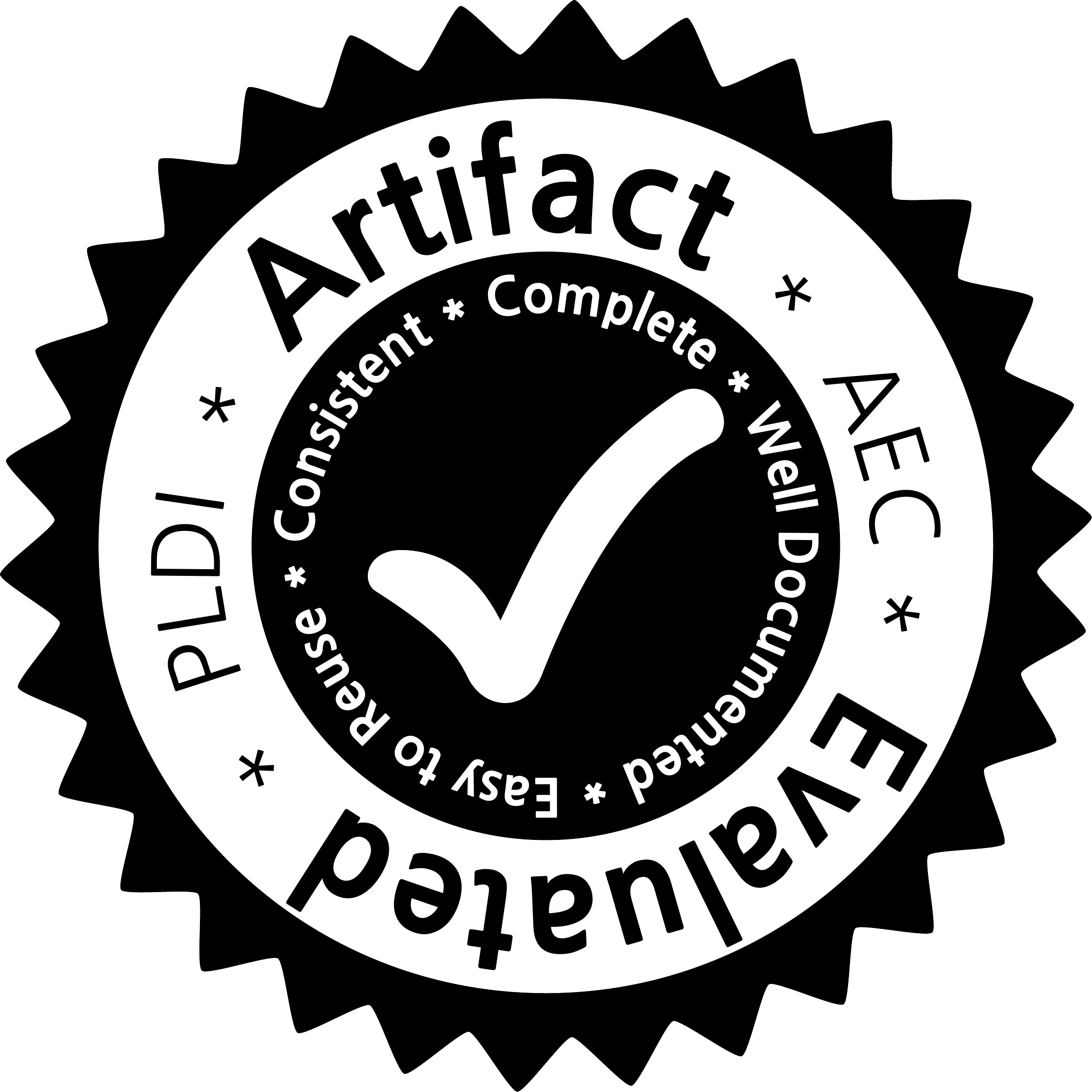}}}
\SetWatermarkAngle{0}

\newif\ifpreprint
\preprintfalse

\newcommand\thetitle{Relatively Complete Counterexamples for Higher-Order Programs}
\ifxetex
   \usepackage{fontspec}
   \usepackage{pifont}
   \newcommand\yay{\ding{51}}
   \newcommand\nay{\ding{55}}
   \usepackage[colorlinks=false,
     draft=false,
     linkcolor=blue,
     citecolor=blue,
     urlcolor=blue, 
     pdftitle={\thetitle}]{hyperref}
\else
   \usepackage[breaklinks=true]{hyperref}
   \usepackage{unixode}
   \usepackage[T1]{fontenc}
   \usepackage[scaled=0.77]{DejaVuSansMono}
   \usepackage{bbding}
   \usepackage{setspace}
   \newcommand\yay{\scalebox{.85}{\Checkmark}}
   \newcommand\nay{\scalebox{.85}{\XSolidBrush}}
\fi

\begin{document}

\paperdoi{2737924.2737971}
\conferenceinfo{PLDI'15}{June 13--17, 2015, Portland, OR, USA}
\CopyrightYear{2015}
\crdata{978-1-4503-3468-6/15/06}
\exclusivelicense

\title{\thetitle}

\authorinfo{Phúc C. Nguy\~{\^e}n}
           {University of Maryland, USA}
           {pcn@cs.umd.edu}
\authorinfo{David Van Horn}
           {University of Maryland, USA}
           {dvanhorn@cs.umd.edu}      

\maketitle 



\setlength{\bibsep}{1pt}

\newtheorem{theorem}{Theorem}
\newtheorem{lemma}{Lemma}
\newtheorem{definition}{Definition}

\definecolor{gray}{rgb}{0.9,0.9,0.9}
\definecolor{red}{rgb}{1,0,0}
\newcommand{\graybox}[1]{\mbox{\setlength{\fboxsep}{0.5pt}%
    \colorbox{gray}{$#1$}}}

\newcommand{\fixme}[1]{{\color{red}{[FIXME: #1]}}}
\newcommand{\todo}[1]{{\color{red}{[TODO: #1]}}}

\newcommand{\ma}[1]{\ensuremath{#1}\xspace}

\newcommand\fakeparagraph[1]{\vspace{-1em}\subsubsection*{#1}}
\newcommand\eval{\mathit{eval}}
\newcommand\aval{\smash[l]{\widehat{\mathit{aval}}}}

\newcommand\refines{\sqsubseteq}
\newcommand\abstracts{\sqsupseteq}

\newcommand\asblm[3]{\ablm{#1}{#2}{#3}{\slambda}{#3}}

\newcommand\toplevel{\ensuremath{\dagger}}
\newcommand\allocname{\ensuremath{\mathsf{alloc}}}

\renewcommand{\vec}[1]{\overrightarrow{#1}}

\newcommand\pow{\ensuremath{\mathcal{P}}}

\newcommand\pluseq{+\!\!=}

\newcommand\ie{\emph{i.e.}}
\newcommand\rstep{{\ensuremath{\mathbf{r}}}}
\newcommand\betav{{\ensuremath{\mathbf{v}}}}
\newcommand\avstep{{\ensuremath{\widehat{\mathbf{v}}}}}
\newcommand\uvstep{{\ensuremath{\avstep\vstep}}}
\newcommand\uvmstep{{\ensuremath{\avstep\vstep{\Delta_M}}}}
\newcommand\betaext{{\ensuremath{\beta_{ext}}}}
\newcommand\abetav{{\ensuremath{\widehat{\mathbf{v}}}}}

\newcommand\tlopaque[2]{\bullet_{#2}^{#1}}
\newcommand\topaque[1]{\bullet^{#1}}
\newcommand\opaque{\bullet}

\newcommand{\dotcup}{\mathrel{\ensuremath{\mathaccent\cdot\cup}}}

\newcommand\with[2]{{#1}/{#2}}
\newcommand\remcon[2]{\with{#1}{#2}}
\newcommand\demonic[2]{\textsc{demonic}(#1,#2)}
\newcommand{\closed}[1]{\dot{#1}}

\newcommand\sclos[2]{\langle{#1},{#2}\rangle}

\newcommand\mans{\ensuremath{A}}
\newcommand\manso{\ensuremath{A'}}
\newcommand\mtyp{\ensuremath{T}}
\newcommand\mbasetyp{\ensuremath{B}}
\newcommand\menv{\ensuremath{\Gamma}}
\newcommand\mprg{\ma{p}}
\newcommand\mprgo{\ensuremath{q}}
\newcommand\mexp{\ensuremath{E}}
\newcommand\maexp{\ensuremath{f}} 
\newcommand\mexpo{\ensuremath{E'}}
\newcommand\mexpoo{\ensuremath{E''}}
\newcommand\mvexp{{\ensuremath{\vec{\mexp}}}}
\newcommand\mvexpo{{\ensuremath{\vec{\mexpo}}}}
\newcommand\mvmod{{\ensuremath{\vec{\mmod}}}}
\newcommand\mvmodo{{\ensuremath{\vec{\mmodo}}}}
\newcommand\mvmodvar{{\ensuremath{\vec{\mmodvar}}}}
\newcommand\mmod{\ensuremath{m}}
\newcommand\mmodo{\ensuremath{m'}}
\newcommand\mmodvar{\ensuremath{f}}
\newcommand\mmodvaro{\ensuremath{g}}
\newcommand\mmodvaroo{\ensuremath{h}}
\newcommand\mcon{\ensuremath{c}}
\newcommand\mcono{\ensuremath{d}}
\newcommand\mbase{\ensuremath{b}}
\newcommand\mpreval{\ensuremath{u}}
\newcommand\mprevalo{\ensuremath{u'}}
\newcommand\mval{\ensuremath{V}\xspace}
\newcommand\mvalo{\ensuremath{V'}} 
\newcommand\mvaloo{\ensuremath{V''}}
\newcommand\mvalset{\ensuremath{\vec\mval}}
\newcommand\maval{\ensuremath{V}} 
\newcommand\mvar{\ensuremath{X}}
\newcommand\mvaro{\ensuremath{Y}}
\newcommand\mvaroo{\ensuremath{Z}}
\newcommand\mnum{\ensuremath{n}}
\newcommand\mnumo{\ensuremath{n'}}
\newcommand\mlam{\ensuremath{l}}
\newcommand\mfun{\ensuremath{w}}
\newcommand\mkont{\ensuremath{\kappa}}
\newcommand\mclos{\ensuremath{D}}
\newcommand\msbl{\ensuremath{S}}
\newcommand\maddr{\ensuremath{a}}
\newcommand\maddro{\ensuremath{b}}
\newcommand\maddroo{\ensuremath{a'}}
\newcommand\maddrtc{\maddro}
\newcommand\mden{\ensuremath{d}}
\newcommand\msto{\ensuremath{\Sigma}}
\newcommand\mstoo{\ensuremath{\Sigma'}}
\newcommand\mstooo{\ensuremath{\Sigma''}}
\newcommand\masto{\ensuremath{\hat\Sigma}}
\newcommand\mastoo{\ensuremath{\hat{\Sigma'}}}
\newcommand\mblm{\ensuremath{B}}
\newcommand\mstate{\ensuremath{\varsigma}}
\newcommand\mastate{\ensuremath{\hat\varsigma}}
\newcommand\mop{O}
\newcommand\moppred{O_?}
\newcommand\mopone{O_1}
\newcommand\moptwo{O_2}
\newcommand\mlab{\ensuremath{L}}
\newcommand\mlabo{\ensuremath{L'}}
\newcommand\mlaboo{\ensuremath{L''}}
\newcommand\mctx{\ensuremath{\mathcal{E}}}
\newcommand\mctxo{\ensuremath{\mathcal{E'}}}
\newcommand\mvval{\ensuremath{\vec\mval}}
\newcommand\mstorable{\ensuremath{S}}
\newcommand\mstorableo{\ensuremath{S'}}
\newcommand\mpred{\ensuremath{P}}
\newcommand\mpredo{\ensuremath{P'}}
\newcommand\mF{\ensuremath{F}}

\newcommand\proves[3]{{{#1}\vdash{#2}:{#3}\mbox{\,\yay}}}
\newcommand\refutes[3]{{#1}\vdash{#2}:{#3}\mbox{\,\nay}}
\newcommand\ambig[3]{{#1}\vdash{#2}:{#3}\,\mbox{\bf ?}} 
\newcommand\extProves[3]{{{#1}{\;\vdash_{S}\;}{#2}:{#3}\mbox{\,\yay}}}
\newcommand\extRefutes[3]{{{#1}{\;\vdash_{S}\;}{#2}:{#3}\mbox{\,\nay}}}
\newcommand\extAmbig[3]{{{#1}{\;\vdash_{S}\;}{#2}:{#3}\mbox{\bf ?}}}
\newcommand\tran[1]{\{\!\!\{#1\}\!\!\}}
\newcommand\lab[2]{lab_{#1}[\![{#2}]\!]}
\newcommand\fv[1]{FV[\![#1]\!]}
\newcommand\build[2]{build_{#1}[\![{#2}]\!]}
\newcommand\app[2]{app?_{#1}[\![{#2}]\!]}

\newcommand\absval{\with\opaque\mconset}

\newcommand\metapp[3]{\metappname({#1},{#2},{#3})}
\newcommand\metappname{\textsc{app}}

\newcommand\MSet[1]{\ensuremath{\mathsf{#1}}}
\newcommand\Mmodvar{\MSet{MVar}}
\newcommand\Mvar{\MSet{Var}}
\newcommand\Mval{\MSet{Val}}
\newcommand\Mcon{\MSet{Con}}
\newcommand\Mprg{\MSet{Prog}}
\newcommand\Mlam{\MSet{Lam}}
\newcommand\Mfun{\MSet{Wrap}}
\newcommand\Mblm{\MSet{Blame}}
\newcommand\Mexp{\MSet{Expr}}

\newcommand\confont[1]{\ensuremath{\mathsf{#1}}}

\newcommand\conarrow{\ensuremath\mbox{\tt →}}
\newcommand\slambda{{\tt λ}}
\newcommand\lparen{\mbox{\texttt{(}}}
\newcommand\rparen{\mbox{\texttt{)}}}

\newcommand\sbegin[2]{#1;\ #2}
\newcommand\sreclamnp[3]{\ensuremath{\slambda_{#2} #1.#3}}
\newcommand\slamnp[2]{\sreclamnp{#1}{\relax}{#2}}
\newcommand\smod[3]{\lparen\ensuremath{\syntax{module}\:#1\:#2\:#3}\rparen}
\newcommand\stlam[3]{\slamnp{#1\!:\!#2}{#3}}
\newcommand\slam[2]{\slamnp{#1}{#2}}
\newcommand\slamp[2]{\lparen\slam{#1}{#2}\rparen}
\newcommand\slamc[3]{\ensuremath{#2\,\conarrow\, \slam{#1}#3}}
\newcommand\sreclam[3]{(\sreclamnp{#1}{#2}{#3})}
\newcommand\strec[3]{\srec{{#1}\!:\!{#2}}{#3}}
\newcommand\srec[2]{\mu{#1}.{#2}}
\newcommand\sapp[2]{\ensuremath{#1\:#2}}
\newcommand\spapp[2]{\ensuremath{\lparen#1\:#2\rparen}}
\newcommand\sapptwo[3]{\ensuremath{(#1\:#2\:#3)}}
\newcommand\sif[3]{\mathsf{if}\:#1\:#2\:#3}
\newcommand\szerop[1]{\mathsf{zero?}(#1)}
\newcommand\sint{\ensuremath{\syntax{int}}}
\newcommand\sany{\ensuremath{\syntax{any}}}
\newcommand\sarr[2]{\ensuremath{#1\,\conarrow\,  #2}}
\newcommand\sdep[3]{\ensuremath{#1\,\conarrow\,\slambda#2.#3}}
\newcommand\stdep[4]{\sdep{#1}{#2\!:\!#3}{#4}}
\newcommand\spred[1]{\ensuremath{\liftpred{#1}}}
\newcommand\scons[2]{(\syntax{cons}\:#1\:#2)}
\newcommand\sconsop{\syntax{cons}}
\newcommand{\stabcons}[2]{(\syntax{cons}\:\begin{tabular}[t]{@{}l}{$#1$}\\{${#2})$}\end{tabular}}
\newcommand\szero{\syntax{zero?}}
\newcommand\ssucc{\syntax{add1}}
\newcommand\sdiv{\syntax{div}}
\newcommand\snatp{\syntax{nat?}}
\newcommand\snump{\syntax{num?}}
\newcommand\sintp{\syntax{int?}}
\newcommand\sequalp{\syntax{=}}
\newcommand\splus{\syntax{+}}
\newcommand\soptwo[3]{{#2}({#1},{#3})}
\newcommand\sopone[2]{{#1}({#2})}
\newcommand\sadd[2]{{#1}+{#2}}
\newcommand\smin[2]{{#1}-{#2}}
\newcommand\sfalse{\syntax{false}}
\newcommand\strue{\syntax{true}}
\newcommand\sboolp{\syntax{bool?}}
\newcommand\sempty{\syntax{empty}}
\newcommand\semptyp{\syntax{empty?}}
\newcommand\scar{\syntax{car}}
\newcommand\scdr{\syntax{cdr}}
\newcommand\sprocp{\syntax{proc?}}
\newcommand\sfalsep{\ensuremath{\mbox{\tt{false?}}}\xspace}
\newcommand\struep{\syntax{true?}}
\newcommand\sconsp{\syntax{cons?}}
\newcommand\sdepp{\syntax{dep?}}
\newcommand\sflatp{\syntax{flat?}}
\newcommand\slistp{\syntax{list?}}
\newcommand\srt[4]{\lparen\syntax{rt}_{\striple{#1}{#2}{#3}}\ #4\rparen}
\newcommand\sblur[4]{\lparen\syntax{blur}_{\striple{#1}{#2}{#3}}\ #4\rparen}
\newcommand\sapprox[2]{#1\ \oplus\ #2}
\newcommand\sref[1]{!#1}
\newcommand\sfin[2]{\ensuremath{\syntax{case}^{#1}\ {#2}}}
\newcommand\shavoc[1]{\syntax{havoc}[\![{#1}]\!]}
\newcommand\smodel[1]{model[\![{#1}]\!]}
\newcommand\sfun[2]{\syntax{fun}\ {#1} \rightarrow {#2}}
\newcommand\slab{\syntax{L}}
\newcommand\slabo{\syntax{L'}}
\newcommand\sboxx{\syntax{box}}
\newcommand\sunbox{\syntax{unbox}}
\newcommand\ssetbox{\syntax{set-box\syntax{!}}}
\newcommand\sboxp{\syntax{box?}}

\newcommand\sext[3]{#1[#2 \mapsto #3]}
\newcommand\sextt[5]{#1[#2 \mapsto #3,#4 \mapsto #5]} 
\newcommand\sexttt[7]{#1[#2 \mapsto #3,#4 \mapsto #5,#6 \mapsto #7]}
\newcommand\sextttt[9]{#1[#2 \mapsto #3,#4 \mapsto #5,#6 \mapsto #7, #8 \mapsto #9]}
\newcommand\sdom[1]{\ensuremath{\mathit{dom}}(#1)}
\newcommand\refine[3]{\ensuremath{\mathit{refine}}(#1, #2, #3)}

\newcommand\syntax[1]{\ensuremath{\mbox{\tt{#1}}}} 

\newcommand\sand[2]{\sif{#1}{#2}{\sfalse}}
\newcommand\sor[2]{\sif{#1}{\strue}{#2}}

\newcommand\sanyc{\ensuremath{\syntax{any/c}}}
\newcommand\snatc{\ensuremath{\syntax{nat/c}}}
\newcommand\sconsc[2]{\langle#1,\!#2\rangle}
\newcommand\sandc[2]{#1\wedge#2}
\newcommand\sorc[2]{#1\vee#2}
\newcommand\srecc[2]{\mu #1.#2}
\newcommand\sboolc{\syntax{bool/c}}
\newcommand\semptyc{\syntax{empty/c}}
\newcommand\spair[2]{\langle#1,\ \!#2\rangle}
\newcommand\striple[3]{\langle#1,#2,#3\rangle}
\newcommand\squadruple[4]{\langle#1,#2,#3,#4\rangle}
\newcommand\sparen[1]{\syntax{(}{#1}\syntax{)}}

\newcommand\sopc[1]{\spred{\slam\mvar{(\sapp{#1}\mvar)}}}
\newcommand\snotopc[1]{\spred{\slam\mvar{\sapp\neg{(\sapp{#1}\mvar)}}}}

\newcommand\tvar{\ensuremath{A}}
\newcommand\tarr[2]{\ensuremath{#1}\rightarrow{#2}}
\newcommand\tn{\syntax{nat}}
\newcommand\tint{\syntax{int}}

\newcommand\amod[2]{\lparen\ensuremath{\syntax{module}\:#1\:#2}\rparen}
\newcommand\achk[6]{\chk{#1}{#3}{#4}{#5}{#2}}
\newcommand\chk[5]{\ensuremath{\syntax{mon}^{#2,#3}_{#4}\mbox{\tt
      (}}#1, #5\mbox{\tt )}}
\newcommand\achksimple[2]{\ensuremath{\syntax{mon}\mbox{\tt
      (}#1,{#2}\mbox{\tt )}}}
\newcommand\afchksimple[2]{\ensuremath{\syntax{fmon}(#1,{#2})}}
\newcommand\ablm[5]{\ensuremath{\syntax{blame}^{#1}_{#2}}}
\newcommand\simpleblm[2]{\ablm{#1}{#2}\relax\relax\relax}
\newcommand\swrong[2]{\ensuremath{\syntax{err}^{#1}_{#2}}}

\newcommand\fc[3]{\ensuremath{\syntax{fc}_{#2}(#1, #3)}}
\newcommand\assume[2]{\ensuremath{\syntax{assume}\lparen{#1},
    #2\rparen}}
\newcommand\sArr[5]{\ensuremath{(#1 \syntax{\Rightarrow}^{#2,#3}_{#4} #5)}}
\newcommand\simpleArr[2]{\ensuremath{\syntax{Arr}(#1, #2)}}

\newcommand\sblm[2]{\ensuremath{\syntax{blame}^{#1}_{#2}}}

\newcommand\stypbool{\syntax{B}}
\newcommand\stypnum{\syntax{N}}
\newcommand\starr[2]{{#1}\rightarrow{#2}}
\newcommand\stcon[1]{\syntax{con}({#1})}
\newcommand\sflat[1]{\ensuremath{\syntax{flat}(#1)}}

\newcommand\svar[1]{\syntax{#1}}

\newcommand\depbless[7]{
  \ma{\slam{#2}{\achk{#3}{(\sapp{#7}{\achk{#1}{#2}{#5}{#4}{#6}{}})}{#4}{#5}{#6}{}}}}
\newcommand\bless[6]{
  \ensuremath{\chk{(#1\!\dashrightarrow\! #2)}{#3}{#4}{#5}{#6}}}

\newcommand\simpledepbless[4]{
  \ma{\slam{#2}{\achksimple{#3}{(\sapp{#4}{\achksimple{#1}{#2}})}}}}
\newcommand\simplebless[3]{
  \ma{\slam\mvar{\achksimple{#2}{(\sapp{#3}{\achksimple{#1}{\mvar}})}}}}

\newcommand\deltamap[5]{\delta({#1},{#2},{#3})\ni{\spair{#4}{#5}}}
\newcommand\absdeltamap[3]{\absdelta({#1},{#2})\ni{#3}}

\newcommand\alloc[1]{\ensuremath{\mathsf{alloc}(#1)}}

\newcommand\subst[3]{\ensuremath{[#1/#2]#3}}
\newcommand\redrule[1]{\ensuremath{\mbox{\sc{#1}}}}
\newcommand\step{\rightarrow}
\newcommand\vstep{\mathbf{v}}       
\newcommand\cstep{\mathbf{c}}       
\newcommand\sstep{\mathbf{s}}       
\newcommand\acstep{{\widehat{\mathbf{c}}}} 
\newcommand\stdstep{\longmapsto}
\newcommand\multistdstep{\stdstep^\star}
\newcommand\mtenv{\ensuremath{\emptyset}}
\newcommand\initstore{[\maddr_0 \mapsto \kmt]}
\newcommand\unload{\ensuremath{\mathcal{U}}}

\newcommand\close[2][\rho]{\ensuremath{\langle{#2},{#1}\rangle}}
\newcommand\mstep{\ensuremath{\stdstep\quad}}
\newenvironment
    {machine}[2][\relax]
    {\begin{display*}[#1]{#2}{\textwidth}\[
\begin{tabular}{@{}>{$}p{2.1in}<{$}>{$}p{.2in}<{$}>{$}p{2.2in}<{$}>{$}p{1.8in}<{$}}
}
    {\end{tabular}\]\end{display*}}   

\newcommand\plaindelta{\cn\delta}
\newcommand\absdelta{\delta}

\newcommand\liftpred[1]{\sflat{#1}\relax}

\newcommand\projleft[1]{\pi_1{#1}}
\newcommand\projright[1]{\pi_2{#1}}
\newcommand\proj\pi

\newcommand\mcache{\mathcal{CV}}

\newcommand\vcons[2]{({#1},{#2})}

\newcommand\s[4]{\langle{#1},{#2},{#4},{#3}\rangle}
\newcommand\se[1]{\s{#1}\menv\msto\mcont}
\newcommand\sk[1]{\s\mval\menv\msto{#1}}

\newcommand\mcont\kappa
\newcommand\mconto\iota

\newcommand\cont[1]{\textsf{#1}}

\newcommand\ab[1]{\widehat{#1}}
\newcommand\cn[1]{\widetilde{#1}}

\newcommand\kchk[6]{\cont{chk}^{#3,#4}_{#5}({#1},{#2},{#6})}
\newcommand\kchkor[8]{\cont{chk-or}^{#5,#6}_{#7}({#1},{#2},{#3},{#4},{#8})}
\newcommand\kchkcons[8]{\cont{chk-cons}^{#3,#4}_{#5}({#1},{#2},{#6},{#7},{#8})}
\newcommand\kchkconso[8]{\cont{chk-cons}^{#5,#6}_{#7}({#1},{#2},{#3},{#4},{#8})}
\newcommand\kfn[4]{\cont{fn}^{#3}({#1},{#2},{#4})}
\newcommand\kap[4]{\cont{ar}^{#3}({#1},{#2},{#4})}
\newcommand\kif[4]{\cont{if}({#1},{#2},{#3},{#4})}
\newcommand\kopone[3]{\cont{op}^{#2}({#1},{#3})}
\newcommand\koptwo[5]{\cont{opr}^{#4}({#1},{#2},{#3},{#5})}
\newcommand\koptwol[5]{\cont{opl}^{#4}({#1},{#2},{#3},{#5})}
\newcommand\kdem[3]{\cont{begin}({#1},{#2},{#3})}
\newcommand\kmt{\cont{mt}}

\newcommand\mkaddr{k}
\newcommand\mkaddro{i}

\newcommand\lang{$\lambda_{\text{C}}$\xspace}
\newcommand\etal{\emph{et al.}}

\newcommand\leftcurly{\ensuremath{\{}}
\newcommand\rightcurly{\ensuremath{\}}}

\newcommand\mloc{\ensuremath{L}}
\newcommand\amb{\mathit{amb}}

\def\TirName{\textit}

\newif\iftechreport

\newcommand\techrep{the accompanying technical report\xspace}
\newcommand\techrepcite{\techrep~\cite{techreport}}

\newcommand\SCV{\textsf{SCV}\xspace}

\newcommand\redname[1]{[\mathit{#1}]}

\techreporttrue

\begin{abstract}
In this paper, we study the problem of generating inputs to a
higher-order program causing it to error.  We first approach the
problem in the setting of PCF, a typed, core functional language and
contribute the first relatively complete method for constructing
counterexamples for PCF programs.  The method is relatively complete
with respect to a first-order solver over the base types of PCF.  In
practice, this means an SMT solver can be used for the effective,
automated generation of higher-order counterexamples for a large class
of programs.

We achieve this result by employing a novel form of symbolic execution
for higher-order programs.   The remarkable aspect of this symbolic
execution is that even though symbolic higher-order inputs and values
are considered, the path condition remains a first-order formula.  Our
handling of symbolic function application enables the reconstruction
of higher-order counterexamples from this first-order formula.

After establishing our main theoretical results, we sketch how to
apply the approach to untyped, higher-order, stateful languages with
first-class contracts and show how counterexample generation can be
used to detect contract violations in this setting.  To validate our
approach, we implement a tool generating counterexamples for erroneous
modules written in Racket.
\end{abstract}

\category{D.2.4}{Software Engineering}{Software/Program Verification}
\category{D.3.1}{Programming Languages}{Formal Definitions and Theory}

\terms{Verification}

\keywords
Higher-order programs; symbolic execution; contracts

\section{Introduction}

Generating inputs that crash first-order programs is a
well-studied problem in the literature on symbolic
execution~\cite{local:Cadar2006EXE,dvanhorn:Godefroid2005DART}, type
systems~\cite{dvanhorn:Foster2002Flowsensitive}, flow
analysis~\cite{dvanhorn:Xie2005Scalable}, and software model
checking~\cite{dvanhorn:yang:modelcheck}.  However, in the setting of
higher-order languages, those that treat computations as first-class
values, research has largely focused on the verification of programs
without investigating how to effectively report counterexamples as
concrete inputs when verification fails (e.g.,
\citet{dvanhorn:Rondon2008Liquid, dvanhorn:Xu2009Static,
  dvanhorn:Kawaguchi2010Dsolve, dvanhorn:Vytiniotis2013HALO,
  dvanhorn:TobinHochstadt2012Higherorder,
  dvanhorn:Nguyen2014Soft}).

There are, however, a few notable exceptions which tackle the problem
of counterexamples for higher-order programs.  Perhaps the most successful
has been the approach of random testing found in tools such as
\emph{QuickCheck}~\cite{dvanhorn:Claessen2000QuickCheck,
  dvanhorn:Klein2010Random}.  While testing works well, it is not a
complete method and often fails to generate inputs for which a little
symbolic reasoning could go further.  Symbolic execution aims to
overcome this hurdle, but previous approaches to higher-order symbolic
execution can only generate \emph{symbolic} inputs, which are not only
less useful to programmers, but may represent infeasible paths in the
program execution~\cite{dvanhorn:TobinHochstadt2012Higherorder,
  dvanhorn:Nguyen2014Soft}.  Higher-order model checking
\cite{dvanhorn:Kobayashi2013Model} offers a complete decision
procedure for typed, higher-order programs with finite base types, and
can generate inputs for programs with potential errors.
Unfortunately, only first-order inputs are allowed.  This assumption
is reasonable for whole programs, but not suitable for testing
higher-order \emph{components}, which often consume and produce
behavioral values (e.g., functions, objects).
\citet{dvanhorn:DBLP:conf/vmcai/ZhuJ13} give an approach to dependent
type inference for ML that relies on counterexample refinement.  This
approach can be used to generate higher-order counterexamples, however
no measure of completeness is considered.

In this paper, we solve the problem of generating potentially
higher-order inputs to functional programs.  We give the first
relatively complete approach to generating counterexamples for PCF
programs.  Our approach uses a novel form of symbolic execution for
PCF that accumulates a path condition as a symbolic heap.  The
semantics is an adaptation of \citet{dvanhorn:Nguyen2014Soft}, where
the critical technical distinction is our semantics maintains a
\emph{complete} path condition during execution. The key insight of
this work is that although the space of higher-order values is huge,
it is only necessary to search for counterexamples from a subset of
functions of specific shapes. Symbolic function application can be
leveraged to decompose unknown functions to lower-order unknown
values.  By the point at which an error is witnessed, there are
sufficient \emph{first-order} constraints to reconstruct the
potentially higher-order inputs needed to crash the program.  The
completeness of generating counterexamples reduces to the completeness
of solving this first-order constraint, and in this way is relatively
complete~\cite{dvanhorn:Cook1978Soundness}.

Beyond PCF, we show the technique is not dependent on assumptions of
the core PCF model such as type safety and purity.  We sketch how the
approach scales to handle untyped, higher-order, imperative programs.
We also show the approach seamlessly scales to handle first-class
behavioral contracts~\cite{dvanhorn:Findler2002Contracts} by
incorporating existing semantics for contract
monitoring~\cite{dvanhorn:Dimoulas2012Complete} with no further work.
The semantic decomposition of higher-order contracts into lower-order
functions naturally composes with our model of unknown functions to
yield a contract counterexample generator for Contract PCF (CPCF).

\paragraph{Contributions}
We make the following contributions:
\begin{enumerate}
\item We give a novel symbolic execution semantics for PCF
  that gradually refines unknown values and maintains a
  complete path condition.
\item We give a method of integrating a first-order solver to
  simultaneously obtain a precise execution of symbolic programs and
  enable the construction of higher-order counterexamples in case of
  errors.
\item We prove our method of finding counterexamples is sound and
  relatively complete.
\item We discuss extensions to our method to handle untyped,
  higher-order, imperative programs with contracts.
\item We implement our approach as an improvement to a previous contract
  verification system, distinguishing definite program errors
  from potentially false positives.
\end{enumerate}

\paragraph{Outline}

The remainder of the paper is organized as follows.  We first step
through a worked example of a higher-order program that consumes
functional inputs (\S~\ref{sec:examples}).
Stepping through the example illustrates the key ideas of
how the path condition is accumulated as a heap of potentially
symbolic values with refinements and how this heap can be translated
to a first-order formula suitable for an SMT solver.  Generating a
model for the path condition at the point of an error reconstructs
the higher-order input needed to witness the error.
Next, we develop the core model of Symbolic PCF (\S~\ref{sec:spcf}) as
a heap-based reduction semantics.  We prove that the semantics is
sound and relatively complete, our main theoretical contribution.  We
then show how to scale the approach beyond PCF to untyped,
higher-order, imperative languages with contracts
(\S~\ref{sec:extensions}).  We use these extensions as the basis of a
tool for finding contract violations in Racket code to validate our
approach (\S~\ref{sec:eval}).  Finally, we relate our work to the
literature (\S~\ref{sec:related}) and conclude (\S~\ref{sec:conclusion}).

\section{Worked examples}
\label{sec:examples}
We illustrate our idea using an \emph{incomplete} OCaml program.  The
basic idea is that we give a semantics to incomplete programs using a
heap of refinements that constrain all possible completions of the
program.  When an error is reached, the heap is given to an SMT
solver, which constructs a model that represents a counterexample.

As our example we use a function \texttt{f} that takes as its
arguments a function \texttt{g} and a number \texttt{n} and performs a
division whose denominator involves the application of \texttt{g} to
\texttt{n}.  We write $\opaque^\mtyp$ to denote an unknown value of
the appropriate type (and omit the type when it is clear from
context).  This example, though contrived, is small and conveys the
heart of our method.
\begin{alltt}
  let f (g : int \(\rightarrow\) int) (n : int) : int =
     1 / (100 - (g n)) 
  in
    (\(\opaque\) f)
\end{alltt}
Now let us consider the possible errors that can arise from running
this code for any interpretation of the unknown value.

Although the application of unknown function $\opaque$
is an arbitrary computation that can result in any error,
we restrict our attention to possible errors stemming from misbehavior
of the visible part of the above code and assume function $\opaque$
is bug-free.
Through symbolic execution and incremental refinement of unknown values,
we reveal one implementation of $\opaque$
that triggers a division error in \texttt{f}'s implementation.
\begin{alltt}
          Error: Division_by_zero
          Breaking context:
           \(\opaque\) = fun f \(\rightarrow\) (f (fun n \(\rightarrow\) 100) 0)
\end{alltt}
To find a counterexample, we first seek a possible error by running
the program under an extended reduction semantics allowing
\emph{unknown}, or \emph{opaque}, values.  When execution follows
different branches, it remembers assumptions associated with each
path, and opaque values become partially \emph{known}, or
\emph{transparent}.  To keep track of incremental refinements
throughout execution, we allocate all values in a heap and maintain an
upper bound to the behavior of each unknown value.

The semantics takes the form of a reduction relation on pairs of
expressions and heaps, written $\spair\mexp\msto \stdstep
\spair\mexpo\mstoo$.  In our example, the first step of computation is
to allocate a fresh location to hold the unknown function being
applied.
\newcommand\mytype{\relax}
\[
\begin{array}{lcl}
\spair{\syntax{(\(\opaque^{\mytype}\) f)}}{\emptyset}
&
\stdstep
&
\spair{\syntax{(\(\slab\sb{1}\) f)}}{[\slab\sb{1} \mapsto \opaque^{\mytype}]} 
\end{array}
\]
Allocating values in the heap this way gives us a means to refine
values and to communicate these refinements to later parts of the
computation.

At this point, we can partially solve for the unknown value.  Since it
is applied to \texttt{f}, it must be a function of one argument.  But
how can we solve for the body of the function?  The key observation is
that while many possible solutions for the function body may exist, if
the function can reach an error state, then it can reach that error
state by immediately applying the input to some arguments,
\emph{without loss of generality}.  Since the input function takes two
arguments, we can partially solve for the body of the function as
``apply the input to two unknown values.''  By allocating these two
unknowns and refining \syntax{f}, we arrive at the state:
\[
\begin{array}{r@{}l}
\langle\syntax{(f \slab\(\sb{2}\) \slab\(\sb{3}\))}, 
[ & \slab\sb{1} \mapsto \sfun{\syntax{f}}{\syntax{(f \(\slab\sb{2}\) \(\slab\sb{3}\))}},\\
  & \slab\sb{2} \mapsto \topaque{\tarr\tint\tint},\\
  & \slab\sb{3} \mapsto \topaque\tint]
\end{array}
\]

The program then executes \texttt{f}'s body, substituting \texttt{g}
with $\slab_2$ and \texttt{n} with $\slab_3$.  The next sub-expression
to reduce is \syntax{(g n)}, which is \syntax{($\slab_2$ $\slab_3$)}
after substitution, which is yet another unknown function application,
so the next step is to partially solve for $\slab_2$.  Unlike in the
higher-order case, there is no interaction with the input value that
needs to be considered (since it is not behavioral), so the function
can simply return a new, unknown output, $\slab_4$, giving us the
following transition:
\[
\begin{array}{cr@{}l}
& \langle\syntax{(\slab\(\sb{2}\) \slab\(\sb{3}\))}, 
 [ & \slab\sb{1} \mapsto \sfun{\syntax{f}}{\syntax{(f \(\slab\sb{2}\) \(\slab\sb{3}\))}},\\
&  & \slab\sb{2} \mapsto \topaque{\tarr\tint\tint},\\
&  & \slab\sb{3} \mapsto \topaque\tint]\rangle\\
\stdstep & \langle\slab_4, [ & \slab\sb{1} \mapsto \sfun{\syntax{f}}{\syntax{(f \(\slab\sb{2}\) \(\slab\sb{3}\))}},\\
  &  & \slab\sb{2} \mapsto \sfun{\syntax{n}}{\slab\sb{4}},\\
 &  & \slab\sb{3} \mapsto \topaque\tint,\\
 &  & \slab\sb{4} \mapsto \topaque\tint]\rangle
\end{array}
\]
At this point, we need to compute \syntax{100 - $\slab_4$},
i.e. subtract an unknown integer from 100.  The solution is simple, we
extend the primitive arithmetic operations to produce new unknown
values and annotate the unknown result with a predicate to embed the
knowledge that it is equal to \syntax{100 - $\slab_4$}:
\[
\begin{array}{cr@{}l}
& \langle\syntax{100 - \(\slab_4\)}, [ & \slab\sb{1} \mapsto \sfun{\syntax{f}}{\syntax{(f \(\slab\sb{2}\) \(\slab\sb{3}\))}},\\
  &  & \slab\sb{2} \mapsto \sfun{\syntax{n}}{\slab\sb{4}},\\
 &  & \slab\sb{3} \mapsto \topaque\tint,\\
 &  & \slab\sb{4} \mapsto \topaque\tint]\rangle\\
\stdstep & 
\langle\slab_5, [ & \slab\sb{1} \mapsto \sfun{\syntax{f}}{\syntax{(f \(\slab\sb{2}\) \(\slab\sb{3}\))}},\\
  &  & \slab\sb{2} \mapsto \sfun{\syntax{n}}{\slab\sb{4}},\\
 &  & \slab\sb{3} \mapsto \topaque\tint,\\
 &  & \slab\sb{4} \mapsto \topaque\tint,\\
 &  & \slab\sb{5} \mapsto \topaque{\tint, \sfun{\syntax{x}}{\syntax{x=(100-\(\slab\sb{4}\))}}}]\rangle
\end{array}
\]
We finally arrive at the point of computing \syntax{1 /
  \(\slab\sb{5}\)}.  At this point the semantics branches
non-deterministically since \(\slab\sb{5}\) may represent a zero or
non-zero value.  In the case of an error, we refine \(\slab\sb{5}\)
to be zero, giving us the final state:
\[
\begin{array}{r@{}l}
\langle\syntax{error}, [ & \slab\sb{1} \mapsto \sfun{\syntax{f}}{\syntax{(f \(\slab\sb{2}\) \(\slab\sb{3}\))}},\\
  & \slab\sb{2} \mapsto \sfun{\syntax{n}}{\slab\sb{4}},\\
  & \slab\sb{3} \mapsto \topaque\tint,\\
  & \slab\sb{4} \mapsto \topaque\tint,\\
  & \slab\sb{5} \mapsto \topaque{\tint, \sfun{\syntax{x}}{\syntax{x=0}}, \sfun{\syntax{x}}{\syntax{x=(100-\(\slab\sb{4}\))}}}]\rangle
\end{array}
\]

At this point, the program has reached an error state and has
accumulated a heap of invariants that constrain the unknown values.
But notice that since functions have been partially solved for as
they've been applied, there are only first-order unknowns in the heap.
At this point, translation of refinements on integers into first-order
assertions is straightforward:
\begin{alltt}
            (declare-const \(\slab\sb{3}\) Int)
            (declare-const \(\slab\sb{4}\) Int)
            (declare-const \(\slab\sb{5}\) Int)
            (assert (= \(\slab\sb{5}\) (- 100 \(\slab\sb{4}\))))
            (assert (= 0 \(\slab\sb{5}\)))
\end{alltt}
A solver such as Z3~\cite{dvanhorn:DeMoura2008Z3} can easily solve
such constraints and yield ($\slab_3$ = \texttt{0}, $\slab_4$ =
\texttt{100}, $\slab_5$ = \texttt{0}) as a model.  We then plug these
values into the current heap and straightforwardly obtain the
counterexample shown at the start.

In summary, we use execution to incrementally construct the shape of
each function, query a first-order solver for a model for base values,
and combine these first-order values to construct a higher-order
counterexample.

\section{Formal model with Symbolic PCF}
\label{sec:spcf}

This section presents a reduction semantics illustrating the core of
our approach.  Symbolic PCF
(SPCF)~\cite{dvanhorn:TobinHochstadt2012Higherorder} extends the PCF
language~\cite{dvanhorn:Scott1993Typetheoretical}
with \emph{incomplete} programs containing \emph{symbolic values} that
can be higher-order.  

We present the language's syntax and semantics, describe its
integration with an external solver, and show how the semantics
enables the generation of a counterexample when an error occurs.
Finally, we prove that our counterexample construction is sound and
complete relative to the underlying solver.  The key technical
challenge in designing such a semantics is to make sure not to
over-constrain unknowns, which would be unsound, while also not
under-constraining unknowns, which would be incomplete.

\subsection{Syntax of SPCF}

Figure~\ref{fig:spcf-syntax} presents the syntax of SPCF.
We write $\vec\mexp$ to mean a sequence of expressions
and treat it as a set where convenient.
The language is simply typed with typical expression forms for conditionals,
applications, primitive applications, recursion,
and values such as natural numbers and lambdas.
The evaluation context $\mctx$ is standard for a call-by-value semantics.
We highlight non-standard forms in gray.
The key extension of SPCF compared to PCF is the notion of \emph{symbolic},
or \emph{opaque} values.
We write $\topaque\mtyp$ to mean an unknown
but fixed and syntactically closed value\footnote{For example, $\opaque$ does not approximate ($\slam{\svar{x}}{\svar{y}}$)}
of type $\mtyp$.
The system automatically annotates each opaque value
with a unique label to identify its source location.
It also uniquely labels each source location that could have a
potential run-time failure.  In SPCF, such failures can only occur
with the application of partial, primitive operations.

When evaluating an SPCF expression, we allocate all values and
maintain a heap to keep track of their constraints.  When execution
proceeds through conditional branches and primitive operations, we
refine the heap at appropriate locations with stronger assumptions
taken at each branch.  As figure~\ref{fig:spcf-syntax} shows, a heap
is a finite function mapping each location $\mlab$ to a stored value
$\mstorable$ as an upper bound of the value's run-time behavior.  A
stored value $\mstorable$ is similar to a syntactic value, but a
stored unknown value can be further refined by arbitrary program
predicates.  For example,
$\topaque{\{\tn,\ \slam{\svar{x}}{\sapp{\syntax{even?}}{\svar{x}}}\}}$
denotes an unknown even natural number.  

In addition, we use $\sfin\mtyp{\vec{[\mlab \mapsto \mlab]}}$ to
denote a mapping approximating an unknown function of type
$\tarr\tn\mtyp$.  We clarify the role of this construct later when
discussing the semantics of applying opaque functions, but the
intuition is that this form is used to constrain unknown functions (of
base type input) to always produce the same result when given the same
input; it is critical for achieving completeness and is not present in
the original SCPCF semantics of
\citet{dvanhorn:TobinHochstadt2012Higherorder}.

Syntax for answers $\mans$ is internal and unavailable to programmers.
An answer is either a location $\mlab^\mtyp$ pointing to a value of
type $\mtyp$ on the heap, or an error message $\swrong\mlab\mop$
\emph{blaming} source location $\mlab$ for violating primitive
$\mop$'s precondition.  A source location in an error message is not
just for precise blaming, but is important in defining what it means
to have a sound symbolic execution, as we will discuss in detail in
section~\ref{sec:spcf-semantics}.

We omit straightforward type-checking rules for SPCF
and assume all considered programs are well-typed.
In addition, we omit showing types and labels for constructs
such as locations and lambdas when they are irrelevant
or clear from context.

In the following, we use the term \emph{unknown program portion}
to refer to all unknown values in the original (incomplete) program,
and \emph{known program portion} to refer to the rest of it.

\begin{figure}[t!]
\[
\begin{array}{l@{\ \,}l@{\;}cl}
{ \mbox{Expressions}} & { \mexp} & ::= & { \graybox\mans\ |\ \mval\ |\ \mvar\ 
          |\ \sif{\!\mexp}{\!\mexp}{\!\mexp}\ |\ \sapp{\!\mexp}{\!\mexp}\ |\ {\sapp\mop{\!\vec\mexp}}\ ^{\mlab}} \\[.8mm]
{ \mbox{Contexts}} & { \mctx} & ::= & { [\ ]\ |\ \sif\mctx\mexp\mexp\ |\
             \sapp\mctx\mexp |\ \sapp\mlab\mctx\ |\ \sapp\mop{\vec\mlab}\mctx{\vec\mexp}} \\[.8mm]
{ \mbox{Values}} & { \mval} & ::= & { \graybox{\tlopaque\mtyp\mlab}\ |\ \slam{\mvar:\mtyp}\mexp\ |\ \mnum}\\[.8mm]
{ \mbox{Answers}} & { \mans} & ::= & { \mlab^\mtyp\ |\ \swrong\mlab\mop}\\[.8mm]
{ \mbox{Operations}} & { \mop} & ::= & { \szero\ |\ \ssucc\ |\ \sdiv\ |\ \ldots}\\[.8mm]
{ \mbox{Predicates}} & { \mpred} & ::= & { \slam{\mvar:\mtyp}\mexp}\\[.8mm]
{ \mbox{Types}} & { \mtyp} & ::= & { \tn\ |\ \tarr{\mtyp}{\mtyp}}\\[.8mm]
{ \mbox{Heaps}} & { \msto} & ::= & { \emptyset\ |\ \msto,\mlab\ \mapsto\ \mstorable}\\[.8mm]
{ \mbox{Storeables}} & {\mstorable} & ::= & {\graybox{\topaque{\{\mtyp \vec\mpred \}}}\
          |\ \slam{\mvar\!:\!\mtyp}\mexp\ |\ \mnum\ |\ \graybox{\sfin\mtyp{\!\vec{\mlab\!\mapsto\!\mlab}}}}\\[.8mm]
{ \mbox{Variables}} & { \mvar,\mlab} & \in & \mathit{identifier}
\end{array}
\]
\caption{Syntax of SPCF}
\label{fig:spcf-syntax}
\end{figure}

\subsection{Semantics of SPCF}
\label{sec:spcf-semantics}

We present the semantics of SPCF as a relation between
states of the form $\spair\mexp\msto$.
Key extensions to the straightforward concrete semantics include
generalization of primitives to operate on symbolic values
and reduction rules for opaque applications.
Intuitively, reduction on abstract states approximates reduction on concrete states,
accounting for all possible instantiations of symbolic values.
Figure~\ref{fig:spcf-semantics} presents the reduction semantics of SPCF.
The semantics is also mechanized as a Redex model and available online.\footnote{\url{https://github.com/philnguyen/soft-contract/tree/pldi-2015/soft-contract/ce-redex}}

Each value is allocated on the heap and reduces to a location
as shown in rules \emph{Opq} and \emph{Conc}.
Because an opaque value stands for an arbitrary but fixed and closed value,
we reuse a location if it has been previously allocated.

Rule \emph{Prim} shows the reduction of a primitive application.
We use $\delta$ to relate primitive operators and values to results.
Typically, $\delta$ is a function, but here it is a relation because
primitive operations may behave non-deterministically on unknown values.
In addition, the relation includes a heap to remember assumptions
in each taken branch.
Rules for conditionals are straightforward,
except we also rely on $\delta$ to determine the truth of the value branched on
instead of replicating the logic.
We use \texttt{0} to indicate falsehood and any non-zero number for truth (as in PCF).
Application of a $\lambda$-abstraction follows standard $\beta$-reduction.

Application of an unknown value to a function argument results in a
range of possibilities to consider.  This space, however, can be
partitioned into a few cases.
First, the unknown program portion can have bugs of its own regardless
of the argument, but our concern is only to find bugs in the known
program portion so the possibility of these errors is ignored.
Second, the function argument escapes into an unknown context and can
be invoked in an arbitrary way.  However, any invocation triggering an
error can be reduced to a chain of function applications.
Alternatively, the unknown function may not explore its argument's
behavior directly during the execution of its body, but delay that in
a returned closure.
Finally, the unknown function may completely ignore its argument and
fail to reveal any hidden bug, allowing the program to proceed to
other parts.
These four cases result in specific shapes a function can have.
Therefore, upon opaque function application,
we refine the opaque function's shape accordingly.

Consider this example:
\begin{alltt}
      (\(\tlopaque{\tarr{(\tarr\tn\tn)}\mtyp}{\slab\sb{1}}\)) (\(\slam{\svar{x}:\tn}\syntax{(/} \syntax{1} \svar{x)}\sp{\slab}\))
\end{alltt}
and the following possible instantiations of $\slab_1$:
\begin{alltt}
    \(1.\) \(\slam{\syntax{f}}{\syntax{(/ 1 0)}}\)\
    \(3.\) \(\slam{\syntax{f}}{\slam{\syntax{x}}{(\sapp\ssucc{(\sapp{\syntax{f}}{\syntax{x}})})}}\)
    \(2.\) \(\slam{\syntax{f}}{\syntax{(f 0)}}\)\
      \(4.\) \(\slam{\syntax{f}}{\slam{\syntax{x}}{\syntax{42}}}\)
\end{alltt}
Completion (1) raises an error from within the unknown function blaming $\slab_1$ itself,
(2) triggers the division error blaming $\slab$,
(3) delays the exploration of its argument's behavior by returning a closure referencing the argument, and
(4) is a constant function ignoring its argument.
As we are only interested in errors in the known program portion, 
we ignore behavior such as (1).
Rules \emph{AppOpq1}, \emph{AppOpq2}, \emph{AppOpq3} and \emph{AppHavoc}
model the remaining possibilities.

Rule \emph{AppOpq1} shows a simple case where the argument is a first-order value with no behavior.
In this case, we approximate the application's result with a symbolic value of appropriate type,
and refine the opaque function to be of the form $\sfin\mtyp{\vec{[\mlab \mapsto \mlab]}}$
to remember this mapping.
Any future application of this function to an equal argument gives an equal result.

Applying a higher-order opaque function results in multiple distinct possibilities.
Rule \emph{AppOpq2} considers the case where the function ignores its argument
(i.e. it is a constant function).
Any future application of this unknown function gives the same result.
Rule \emph{AppOpq3} considers the case where the unknown context
does not immediately explore its argument's behavior
but delays that work by wrapping the argument inside another function.
The context using this result may or may not reveal a potential error.
Finally, rule \emph{AppHavoc} considers the case where the unknown context
explores its argument's behavior by supplying an unknown value to its argument
and putting the result back into another unknown context.

When the argument is higher-order, we do not use a simple dispatch as in rule \emph{AppOpq1}
because there is no mechanism for comparing functions for equality
(without applying them as in rule \emph{AppHavoc}).

Application rules for mappings are straightforward.
Rule \emph{AppCase1} reuses the result's location for a
previously seen application,
whereas rule \emph{AppCase2} allocates a fresh location for the result
of a newly seen application.

These rules for opaque application collectively model the \emph{demonic} context
in previous work on higher-order symbolic execution~\cite{dvanhorn:TobinHochstadt2012Higherorder},
but they unroll the unknown context incrementally
and remember its shape to enable counterexample construction when execution finishes.

Finally, we define the semantics to be the contextual closure of all
the above reductions (rule \emph{Close}).
Errors halt the program and discard the context (rule \emph{Error}).

\begin{figure*}[t!]
\[
\begin{array}{l@{\ }l@{\ }r}

\spair{\tlopaque\mtyp\mlab}\msto
 \stdstep
 \spair{\mlab^\mtyp}\mstoo
&\mbox{where }\mstoo = \sext\msto\mlab{\tlopaque\mtyp\mlab}\mbox{ if }\mlab \notin \sdom\msto
\mbox{, or }\msto\mbox{ otherwise}
&\redname{Opq}
\\


\spair\mval\msto
 \stdstep
 \spair\mlab{\sext\msto\mlab\mval}
&\mbox{where }\mlab \notin \sdom\msto\mbox{ and }\mval \neq \opaque
&\redname{Conc}
\\

%


\spair{\sif\mlab{\mexp_1}{\mexp_2}}\msto
 \stdstep
 \spair{\mexp_1}\mstoo
&\deltamap\msto\szero\mlab{0}\mstoo
&\redname{IfTrue}
\\

\spair{\sif\mlab{\mexp_1}{\mexp_2}}\msto
 \stdstep
 \spair{\mexp_2}\mstoo
&\deltamap\msto\szero\mlab{1}\mstoo
&\redname{IfFalse}
\\

\spair{(\sapp\mop{\vec\mlab})}\msto
 \stdstep
 \spair\mlabo{\sext\mstoo\mlabo\mval}
&\mbox{if }\deltamap\msto\mop{\vec\mlab}\mval\mstoo \mbox{ and } \mlabo\ \notin\ \sdom\mstoo
&\redname{Prim}
\\

\spair{(\sapp\mlab{\mlab_x})}\msto
 \stdstep
 \spair{\subst{\mlab_x}\mvar\mexp}\msto
&\mbox{if }\sapp\msto{(\mlab)}\ =\ \slam\mvar\mexp
&\redname{AppLam}
\\


\spair{(\sapp\mlab{\mlab_x})}\msto
 \stdstep

&\mbox{if }\sapp\msto{(\mlab)}\ =\ \topaque{\tarr\tn\mtyp}
 \mbox{and }\mlab_a\ \notin\ \sdom\msto
&\redname{AppOpq1}
\\
\quad \spair{\mlab_a}
       {\sextt\msto{\mlab_a}{\topaque\mtyp}\mlab{\sfin\mtyp{[\mlab_x \mapsto \mlab_a]}}}
\\

\spair{(\sapp\mlab{\mlab_x})}\msto
 \stdstep
 \spair{\mlab_a}
       {\sextt\msto{\mlab_a}{\topaque\mtyp}\mlab{\slam{\svar{x}:{\mtyp'}}{\mlab_a}}}

& \mbox{if }\sapp\msto{(\mlab)}\ =\ \topaque{\tarr{\mtyp'}\mtyp}
       \mbox{, }\mtyp' = \tarr{\mtyp_1}{\mtyp_2}
       \mbox{ and }\mlab_a\ \notin\ \sdom\msto
&\redname{AppOpq2}
\\

\spair{(\sapp\mlab{\mlab_x})}\msto
 \stdstep
 \spair{\subst{\mlab_x}{\svar{x}}\mval}\mstoo 

&\mbox{if }\sapp\msto{(\mlab)}\ =\ \topaque{\tarr{\mtyp'}\mtyp}
 \mbox{, }\mtyp' = \tarr{\mtyp_1}{\mtyp_2}
 \mbox{, }\mtyp = \tarr{\mtyp_3}{\mtyp_4}
&\redname{AppOpq3}\\
&\multicolumn{2}{l}{\mbox{where }\mstoo = \sextt\msto{\mlab}{\slam{\svar{x}:\mtyp'}{\mval}}
                                                     {\mlab_1}{\topaque{\tarr{\mtyp'}\mtyp}}}\\
&\multicolumn{2}{l}{{\mlab_1}\ \notin\ \sdom\msto
                    \mbox{, and }\mval = \slam{\svar{y}:{\mtyp_3}}{(\sapp{(\sapp{\mlab_1}{\svar{x}})}{\svar{y}})}}
\\

\spair{(\sapp\mlab{\mlab_x})}\msto
 \stdstep
&\mbox{if }\sapp\msto{(\mlab)}\ =\ \topaque{\tarr{\mtyp'}\mtyp}
      \mbox{, }\mtyp' = \tarr{\mtyp_1}{\mtyp_2}, 
&\redname{AppHavoc}
\\

\quad
\spair{(\sapp{\mlab_2}{(\sapp{\mlab_x}{\mlab_1})})}
      {\sexttt\msto\mlab\mval
        {\mlab_1}{\topaque{\mtyp_1}}
        {\mlab_2}{\topaque{\tarr{\mtyp_2}\mtyp}}}
&
{\mlab_1},{\mlab_2},{\mlab_a}\ \notin\ \sdom\msto
\mbox{, and }\mval = \slam{{\svar{x}}:\mtyp'}{(\sapp{\mlab_2}{(\sapp{\svar{x}}{\mlab_1})})}
\\

\spair{(\sapp\mlab{\mlab_x})}\msto
 \stdstep
 \spair{\mlab_a}\msto
&\mbox{if }\sapp\msto{(\mlab)}\ =\ \sfin\relax{\ldots\ [\mlab_x\ \mapsto\ \mlab_a]\ \ldots}\relax
&\redname{AppCase1}
\\

\spair{(\sapp\mlab{\mlab_x})}\msto
 \stdstep
&\mbox{if }\sapp\msto{(\mlab)}\ =\ \sfin\relax{[{\mlab_z}\ \mapsto\ {\mlab_b}]\ldots}\relax
\mbox{and }\mlab_x\ \notin\ \{{\mlab_z}\ldots \}
&\redname{AppCase2}
\\

\quad
\spair{\mlab_a}{\sext\msto\mlab{\sfin\relax{[{\mlab_z}\ \mapsto\ {\mlab_b}]\ldots\ [{\mlab_x}\ \mapsto\ {\mlab_a}]}\relax}}
& \mbox{ and }\mlab_a\ \notin\ \sdom\msto
\\

\spair{\mctx[\mexp]}\msto
 \stdstep
 \spair{\mctx[\mexpo]}\mstoo
&\mbox{if }\spair{\mexp}\msto \stdstep \spair\mexpo\mstoo
&\redname{Close}
\\

\spair{\mctx[\swrong\mlab\mop]}\msto
 \stdstep
 \spair{\swrong\mlab\mop}\msto
&\mbox{if }\mctx \neq [\ ]
&\redname{Error}

\end{array}
\]
\caption{Semantics of SPCF}
\label{fig:spcf-semantics}
\end{figure*}

\subsection{Primitive operations}

We rely on relation $\delta$ to interpret primitive operations.
The rules straightforwardly extend standard operators to work on symbolic values.
In particular, division by an unknown denominator non-deterministically
either returns another integer or raises an error.
The relation also remembers appropriate refinements to arguments and results at each branch. 
Figure~\ref{fig:spcf-delta} presents a selection of representative rules for primitive operations
$\szero$ and $\sdiv$.
We abbreviate $\slam\mvar{\syntax{(}\sequalp \mvar \mexp\syntax{)}}$
as $\syntax{(}\sapp{\syntax{\(\equiv\)}}\mexp\syntax{)}$.
Rules for primitive predicates such as $\szero$ utilize a proof relation
between the heap, the value, and a predicate, which we present next.

\begin{figure}
\[
\begin{array}{l@{\ }r}
\deltamap\msto\szero\mlab{\syntax{1}}\msto
&\mbox{if }\proves\msto\mlab\szero\\

\deltamap\msto\szero\mlab{\syntax{0}}\msto
&\mbox{if }\refutes\msto\mlab\szero\\

\multicolumn{2}{l}{
\delta(\msto,\szero,\mlab) \supseteq \{\spair{\syntax{1}}{\sext\msto\mlab{\syntax{0}}}, \spair{\syntax{0}}{\sext\msto\mlab{\neg\szero}}\}
}\\
&\mbox{if }\ambig\msto\mlab\szero\\[2mm]



\multicolumn{2}{l}{\deltamap\msto\sdiv{\mlab_1, \mlab_2}{m/n}\msto}\\
\multicolumn{2}{r}{
\mbox{if }\msto(\mlab_1) = m
\mbox{ and }\msto(\mlab_2) = n, n \neq \syntax{0}}\\
\multicolumn{2}{l}{\deltamap\msto\sdiv{\mlab_1, \mlab_2}{\topaque{\tn,\ \syntax{(\(\equiv\) \(\mlab_1\) / \(\mlab_2\))}}}\mstoo}\\
\multicolumn{2}{r}{
\mbox{if }\msto(\mlab_2) \neq n
\mbox{ and }\deltamap\msto\szero{\mlab_2}{\syntax{0}}\mstoo}\\
\multicolumn{2}{l}{\deltamap\msto\sdiv{\mlab_1, \mlab_2}{\swrong{}\sdiv}\mstoo}\\
\multicolumn{2}{r}{
\mbox{if }\msto(\mlab_2) \neq n
\mbox{ and }\deltamap\msto\szero{\mlab_2}{\syntax{1}}\mstoo}
\end{array}
\]
\caption{Selected Primitive Operations}
\label{fig:spcf-delta}
\end{figure}

\subsection{Proof relation}

We define a proof relation deciding whether a value satisfies a predicate.
We write $\proves\msto\mlab\mpred$ to mean the value at location $\mlab$ definitely
satisfies predicate $\mpred$, which implies that all possible
instantiations of $\mlab$ satisfy $\mpred$.
In the same way, $\refutes\msto\mlab\mpred$ means all instantiations of $\mlab$
definitely fail $\mpred$.
Finally, $\ambig\msto\mlab\mpred$ is a conservative answer when
we cannot draw a conclusion given information from existing refinements
on the heap.

Precision of our execution relies on this proof relation.
(A trivial relation answering ``neither'' for all queries would make
the execution sound though highly imprecise.)
Instead of implementing our own proof system, we rely on an
SMT solver for sophisticated reasoning on base values.

Figure~\ref{fig:heap-translation} shows the translation $\tran{\cdot}$
of run-time constructs into logical formulas.  The translation of a
heap is the conjunction of formulas obtained from each location and its value, and
the translation of each location and value is straightforward.  In particular, a
location pointing to a concrete number translates to the obvious
assertion of equality, and a mapping ($\sfin{}{\vec{\mlab \mapsto
    \mlab}}$) adds constraints asserting that equal inputs imply equal
outputs.  Since outputs of maps may be functions, it might appear as
though we need function equality.  However, we do not need general
equality on functions, but just a specialized equality that can handle
those opaque functions generated by \emph{AppOpq1}, \emph{AppOpq2},
\emph{AppOpq3} and \emph{AppHavoc}.  Equality on similar function forms
proceeds structurally, while equality on different function forms
translate trivially to \syntax{False} (not shown).

Notice that the proof system only needs to handle predicates of simple forms
and not their arbitrary compositions.
We rely on execution itself to break down complex predicates to smaller ones
and take care of issues such as divergence and errors in the predicate itself.
For example, if the proof system can prove that a value satisfies predicate $\mpred$,
it automatically allows the execution to prove that the value also satisfies
$\syntax{(}\slam{\svar{x}}{\syntax{(}\texttt{or}\ \syntax{(}\mpred\ \svar{x}\syntax{)}\ \mexp\syntax{))}}$ for an arbitrarily expression $\mexp$.
By the time we have [$\mlab \mapsto \topaque{\vec\mpred}$],
we can assume all predicates $\vec\mpred$ have terminated with true on $\mlab$.
Further, because many solvers do not support uninterpreted higher-order functions,
we do not assume such a feature, and the translation only produces queries on first-order values.
Nevertheless, the symbolic execution itself can reason about higher-order unknown values.
Handling higher-order functions on the semantics side and
not relying on the theory of uninterpreted functions
also potentially allows the method
to scale to more realistic language features such as side effects.

For each query between heap $\msto$, location $\mlab$ and predicate $\mpred$,
we translate known assumptions from the heap to obtain formula \ensuremath{\phi},
and the relationship ($\mlab$ : $\mpred$) to obtain formula \ensuremath{\psi}.
We then consult the solver to obtain an answer.
As figure~\ref{fig:proof-rules} shows,
validity of \sparen{\ensuremath{\phi} \texttt{\(\Rightarrow\)} \ensuremath{\psi}}
implies that value $\mlab$ definitely satisfies predicate $\mpred$,
and unsatisfiability of \sparen{\ensuremath{\phi} \texttt{\(\wedge\)} \ensuremath{\psi}} means value $\mlab$ definitely refutes $\mpred$.
If neither can be determined, we return the conservative answer.

\begin{figure}
\[
\begin{array}{l}

\tran{\vec{\mlab \mapsto \mstorable}} = \bigwedge{\vec{\tran{\mlab \mapsto \mstorable}}}\\
\tran{\mlab \mapsto n} = \sparen{\mlab\ \syntax{=}\ n}\\
\tran{\mlab \mapsto \topaque{\tn\ \vec\mpred}} = \bigwedge\ \vec{\tran{\mlab : \mpred}}\\
\tran{\mlab \mapsto \sfin{}{\_\ldots[\mlab_1 \mapsto \mlab_2]\ \_\ldots[\mlab_3 \mapsto \mlab_4]\ \_\ldots}}\\
\multicolumn{1}{r}{= \sparen{\bigwedge\ \sparen{\sparen{\mlab_1\ \syntax{=}\ \mlab_3}\ \syntax{\(\Rightarrow\)}\ \sparen{\mlab_2\ \syntax{=}\ \mlab_4}}\ldots}}\\[2mm]
\tran{\mlab : (\slam\mvar{\sapp{\szero\ \mvar}})} = \syntax{(\mlab\ =\ 0)}\\
\tran{\mlab : (\slam\mvar{(= \mvar (+\ \mlab_1\ \mlab_2))})} =
\sparen{\mlab\ \syntax{=}\ \sparen{\mlab_1\ \syntax{+}\ \mlab_2}}\\

\tran{\mlab_1\ \syntax{=}\ \mlab_2}_{\tn} = \sparen{\mlab_1\ \syntax{=}\ \mlab_2}\\
\tran{\mlab_1\ \syntax{=}\ \mlab_2}_{\tarr{\mtyp}{\mtyp'}} = \tran{\msto(\mlab_1)\ \syntax{=}\ \msto(\mlab_2)}\\
\tran{\sparen{\sfin{\mtyp}{[\mlab_1 \mapsto \mlab_2]\ldots}\ \syntax{=}
              \sfin{\mtyp}{[\mlab_3 \mapsto \mlab_4]\ldots}}} =\\
  \multicolumn{1}{r}{
  \sparen{\bigwedge{\tran{(\mlab_1\ \syntax{=}\ \mlab_3) \Rightarrow (\mlab_2\ \syntax{=}\ \mlab_4)}\ldots}}}\\
\tran{\slam\mvar{\mlab_1}\ \syntax{=}\ \slam\mvar{\mlab_2}} = \tran{\mlab_1\ \syntax{=}\ \mlab_2}\\
\tran{\slam\mvar{(\sapp{\mlab_1}{(\sapp{\mvar}{\mlab_2})})}\ \syntax{=}\
      \slam\mvar{(\sapp{\mlab_3}{(\sapp{\mvar}{\mlab_4})})}} =\\
 \multicolumn{1}{r}{
 \sparen{\wedge\ \tran{\mlab_1\ \syntax{=}\ \mlab_3}\ \tran{\mlab_2\ \syntax{=}\ \mlab_4}}}\\
\tran{\slam\mvar{\slam\mvaro{(\sapp{(\sapp{\mlab_1}{\mvar})}{\mvaro})}}\ \syntax{=}
      \slam\mvar{\slam\mvaro{(\sapp{(\sapp{\mlab_2}{\mvar})}{\mvaro})}}} =\\
 \multicolumn{1}{r}{
 \tran{\mlab_1\ \syntax{=}\ \mlab_2}}\\
\end{array}
\]
\caption{Heap translation}
\label{fig:heap-translation}
\end{figure}

\begin{figure}
\begin{mathpar}

\inferrule[Proved]
{\tran\msto \Rightarrow \tran{\mlab : \mpred}\mbox{ is valid}}
{\proves\msto\mlab\mpred}
\ \ \ \ \
\inferrule[Refuted]
{\tran\msto \wedge \tran{\mlab : \mpred}\mbox{ is unsat}}
{\refutes\msto\mlab\mpred}

\inferrule[Ambig]
{\tran\msto \Rightarrow \tran{\mlab : \mpred}\mbox{ is invalid}
 \mbox{ and }\tran\msto \wedge \tran{\mlab : \mpred}\mbox{ is sat}}
{\ambig\msto\mlab\mpred}

\end{mathpar}
\caption{Proof rules}
\label{fig:proof-rules}
\end{figure}

\subsection{Constructing counterexamples}

For each answer reached by evaluation,
the heap contains refinements to symbolic values in order to reach such results.
In particular, refinements on the heap in an error case
describe the condition under which the program goes wrong.

Specifically, at the end of evaluation, refinements on the heap are nearly concrete:
higher-order symbolic values are broken down into a chain of argument deconstruction and mappings,
and first-order symbolic values have precise constraints
that identify the execution path.
Indeed, a model to the first-order constraints on the heap yields a counterexample
to the program.
We simply plug first-order concrete values back into the heap.

The reader may wonder if this process always generates an actual counterexample
witnessing a real program bug (soundness),
and if it always finds counterexample when a bug exists (completeness).
The next section clarifies these points.

\subsection{Soundness and completeness of counterexamples}
\label{sec:spcf-theorems}

We show that our method of finding counterexamples in a higher-order
program is sound and relatively complete.  Soundness means that the
system only gives an actual counterexample triggering a bug (not a
false positive).  Relative completeness means that if the program
actually contains a bug and the underlying solver can answer all
queries on first order data, the system constructs a concrete
counterexample witnessing that bug, even when it involves complex
interactions between higher-order values.

The statements and proofs of soundness and completeness revolve around
a notion of \emph{approximation}, which we first describe before
stating our main theorems.

\paragraph{Approximation Relation}

We define an approximation relation between \emph{concrete} and
\emph{abstract} states.  A concrete state contains no unknown values,
while an abstract state may contain unknowns.  We write
$\spair\mexpo\mstoo \refines \spair\mexp\msto$ to mean
``$\spair\mexp\msto$ \emph{approximates} $\spair\mexpo\mstoo$,'' or
conversely, ``$\spair\mexpo\mstoo$ \emph{instantiates}
$\spair\mexp\msto$,'' where $\spair\mexpo\mstoo$ is a concrete state
and $\spair\mexp\msto$ is an abstract state.  We make two remarks
about the relation before defining it.

First, as discussed in section~\ref{sec:spcf-semantics},
when analyzing an incomplete program, we are only concerned with errors
coming from known code.
Therefore, we parameterize the approximation relation with a set of labels
$\vec\mlab$ denoting application sites from the known program portion.
Figure~\ref{fig:lab} presents the straightforward definition of metafunction $\ensuremath{lab}$
for computing a program's labels identifying application sites.
The function takes a heap to compute labels for intermediate states,
where a function may be referenced indirectly through its location.
For the purpose of analyzing program $\mexp$,
the set of labels is $\lab\emptyset\mexp$.
As an example, expression $\mexp$ below has an instantiation $\mexpo$,
but when analyzing $\mexp$, we are only interested in the potential division
error at $\slab$ and not $\slabo$.
\begin{alltt}
        \(\mexp = \sparen{\sdiv \syntax{1} \sparen{\sapp{\topaque{\tarr\tint\tint}}{\syntax{1}}}}\sp{\slab}\)
        \(\mexpo = \sparen{\sdiv \syntax{1} \sparen{\sapp{\slam{\svar{x}}{\sparen{\sdiv \syntax{1} \svar{x}}\sp{\slabo}}}{\syntax{1}}}}\sp{\slab}\)
\end{alltt}

Second, we enforce that each location in the abstract state unambiguously approximates
one location in the concrete state
by parameterizing the approximation relation with a function $\mF$
mapping each label in the abstract state to one in the concrete state.
For example, we do not want the following concrete state $\spair\mexpo\mstoo$
to instantiate the abstract state $\spair\mexp\msto$,
even though $\topaque\tint$ intuitively approximates each number \syntax{1},
\syntax{2}, and \syntax{3} individually.
\begin{alltt}
\(\spair{\mexp}{\msto}\)  = \(\spair{\sparen{\sif\slab\slab\slab}   }{\{\slab \mapsto \topaque\tint\}}\)
\(\spair{\mexpo}{\mstoo}\) = \(\spair{\sparen{\sif{\slab\sb{1}}{\slab\sb{2}}{\slab\sb{3}}}}{\{\slab\sb{1} \mapsto 1, \slab\sb{2} \mapsto 2, \slab\sb{3} \mapsto 3\}}\)
\end{alltt}
Instead, the following concrete state $\spair\mexpoo\mstooo$ properly instantiates
$\spair\mexp\msto$ with function $\mF$ = $\{\slab \mapsto \slab_1\}$:
\begin{alltt}
        \(\spair\mexpoo\mstooo\) = \(\spair{\sparen{\sif{\slab\sb{1}}{\slab\sb{1}}{\slab\sb{1}}}}{\{\slab\sb{1} \mapsto 1\}}\)
\end{alltt}


Figure~\ref{fig:approx} defines the approximation parameterized by
label set $\vec\mlab$ and function $\mF$.  We present important,
non-structural rules for the approximation relation between heaps,
values, and states.  We omit displaying parameters when they are
unimportant or can be inferred from context.  We defer the full
definition to the appendix of the extended version of this
paper~\cite{dvanhorn:DBLP:journals/corr/NguyenH15}.

Rule \emph{Heap-Ext} states that if a heap approximates a concrete heap,
it approximates any extension of that concrete heap.
This rule is necessary for ignoring irrelevant computations
in instantiation of an opaque function.
Next, rules \emph{Heap-Int}, \emph{Heap-Lam}, \emph{Heap-Opq-1}, and \emph{Heap-Opq-2}
show straightforward extensions to the approximation when the heaps on both sides
are extended.
First, any concrete value of the right type instantiates
the opaque value $\topaque\mtyp$ as long as the instantiating value does not
contain source locations from the known program portion.
Second, refining an abstract value with a predicate known to be satisfied
by the concrete value preserves the approximation relation.
Because a predicate can contain locations,
we substitute labels appropriately as indicated by function $\mF$.
We omit the straightforward definition of this substitution.

Rules \emph{Heap-Case-1} and \emph{Heap-Case-2} establish the approximation between
functions on natural numbers.
First, a fully opaque mapping approximates all functions.
In addition, if there exists an execution trace witnessing
that applying the concrete function yields a value approximated by
an opaque value, then refining the mapping preserves the approximation.

Rule \emph{Loc} states that location $\mlab$ approximates $\mlabo$
if the pair agrees with function $\mF$.

Rule \emph{Err-Opq} reflects our decision of ignoring errors blaming
source locations from unknown code.  Otherwise, rule \emph{Err} says
that an error with a known label approximates another when they are
the same error.

Finally, rule \emph{Opq-App} states that we ignore irrelevant
computation from a concrete function that instantiates an unknown
function.  Specifically, if we can establish that an opaque
application approximates a concrete application by the structural
rule, then the opaque application continues to approximate each
non-answer state reachable from the concrete application.  There are
similar rules for approximation by applying other forms of opaque
functions, which we defer to the appendix of the extended version of
this paper~\cite{dvanhorn:DBLP:journals/corr/NguyenH15}.

\begin{figure}
\[
\begin{array}{rcl}

\lab\msto{{(\sapp\mop\mexp)}^\mlab}
&=&
\{\mlab\} \cup \lab\msto\mexp\\

\lab\msto{\sapp{\mexp_1}{\mexp_2}}
&=&
\lab\msto{\mexp_1} \cup \lab\msto{\mexp_2}\\

\lab\msto{\sif\mexp{\mexp_1}{\mexp_2}}
&=&
\lab\msto\mexp \cup \lab\msto{\mexp_1} \cup \lab\msto{\mexp_2}\\


\lab\msto{\slam\mvar\mexp}
&=&
\lab\msto\mexp\\

\lab\msto\mlab
&=&
\lab\msto{\msto(\mlab)}\\

\lab\msto{\_}&=&\emptyset
\end{array}
\]
\caption{Computing concrete labels}
\label{fig:lab}
\end{figure}

\begin{figure}
\begin{mathpar}


\inferrule[Heap-Empty]
{}
{\emptyset \refines^{\{\}}_{\vec\mlab} \emptyset}

\inferrule[Heap-Ext]
{\mstoo \refines^\mF_{\vec\mlab} \msto}
{\sext\mstoo\mlabo\mstorableo \refines^\mF_{\vec\mlab} \msto}

\inferrule[Heap-Int]
{\mstoo \refines^\mF_{\vec\mlab} \msto}
{\sext\mstoo\mlabo\mnum \refines^{\sext\mF\mlab\mlabo}_{\vec\mlab} \sext\msto\mlab\mnum}

\inferrule[Heap-Lam]
{\mstoo \refines^{\mF}_{\vec\mlab} \msto\\
 \spair\mexpo\mstoo \refines^\mF_{\vec\mlab} \spair\mexp\msto}
{\sext\mstoo\mlabo{\slam\mvar\mexpo} \refines^{\sext\mF\mlab\mlabo}_{\vec\mlab} \sext\msto\mlab{\slam\mvar\mexp}}

\inferrule[Heap-Opq-1]
{\mstoo \refines^{\mF}_{\vec\mlab} \msto\\
 \lab\mstoo\mvalo \cap \vec\mlab = \emptyset}
{\sext\mstoo\mlabo\mvalo \refines^{\sext\mF\mlab\mlabo}_{\vec\mlab} \sext\msto\mlab{\topaque\mtyp}}

\inferrule[Heap-Opq-2]
{\sext\mstoo\mlabo\mvalo \refines^\mF_{\vec\mlab} \sext\msto\mlab{\topaque{\mtyp \mpred\ldots}}\\
 \proves\mstoo\mvalo{\mF({\mpred_1})}}
{\sext\mstoo\mlabo\mvalo \refines^\mF_{\vec\mlab} \sext\msto\mlab{\topaque{\mtyp \mpred\ldots{\mpred_1}}}}

\inferrule[Heap-Case-1]
{\mstoo \refines^{\mF}_{\vec\mlab} \msto\\
 \lab\mstoo\mexpo \cap \vec\mlab = \emptyset}
{\sext\mstoo\mlabo{\slam\mvar\mexpo} \refines^{\sext\mF\mlab\mlabo}_{\vec\mlab} \sext\msto\mlab{\sfin\mtyp{[\ ]}}}

\inferrule[Heap-Case-2]
{\sext\mstooo\mlabo{\slam\mvar\mexpo} \refines^{\mF}_{\vec\mlab} \sext\msto\mlab{\sfin\mtyp{[\ldots]}}\\
 \mF(\mlab_x) = \mlabo_x\\
 \spair{\subst\mvar{\mlabo_x}\mexpo}\mstooo \multistdstep \spair\mvalo\mstoo\\
 \spair\mvalo\mstoo \refines^\mF_{\vec\mlab} \spair\mval\msto}
{\sext\mstoo\mlabo{\slam\mvar\mexpo} \refines^{\sext\mF\mlab\mlabo}_{\vec\mlab} \sext\msto\mlab{\sfin\mtyp{[\ldots\mlab_x \mapsto \mval]}}}


\inferrule[Loc]
{\mF(\mlab) = \mlabo}
{\spair\mlabo\mstoo \refines^\mF_{\vec\mlab} \spair\mlab\msto}

\inferrule[Err-Opq]
{\mlabo \notin \vec\mlab}
{\spair{\swrong\mlabo\mop}\mstoo
 \refines^\mF_{\vec\mlab}
 \spair\mexp\msto}

\inferrule[Err]
{\mlabo \in \vec\mlab}
{\spair{\swrong\mlab\mop}\mstoo
 \refines^\mF_{\vec\mlab}
 \spair{\swrong\mlab\mop}\msto}

\inferrule[Opq-App]
{\mexpo \neq \mans\\
 \lab\mstoo\mexpoo \cap \vec\mlab = \emptyset\\
 \spair{\sapp{\mlabo_f}{\mlabo_x}}\mstooo \refines^\mF_{\vec\mlab} \spair{\sapp{\mlab_f}{\mlab_x}}\msto\\
 \mstooo(\mlabo_f) = \slam\mvar\mexpoo\\
 \msto(\mlab_f) = \topaque{\tarr\mtyp\mtyp'}\\
 \spair{\subst\mvar{\mlabo_x}\mexpoo}\mstooo \multistdstep \spair\mexpo\mstoo
 }
{\spair\mexpo\mstoo
 \refines^\mF_{\vec\mlab}
 \spair{(\sapp{\mlab_f}{\mlab_x})}\msto}

\end{mathpar}
\caption{Approximation}
\label{fig:approx}
\end{figure}


Theorem~\ref{thm:ce-sound} states that every constructed counterexample
from an error case actually reproduces the same error.
Notice that the theorem is conditioned on $\mstoo \refines \msto_2$
and does not imply that all errors in the abstract execution are real.
In particular, if a path is spurious, its final heap has no instantiation.
\begin{theorem}[Soundness of Counterexamples]
\label{thm:ce-sound}\ \\
If $\spair\mexp{\msto_1}\ \multistdstep\ \spair{\swrong\mlab\mop}{\msto_2}$
and $\mstoo \refines \msto_2$,
then $\spair\mexp\mstoo\ \multistdstep\ \spair{\swrong\mlab\mop}\mstooo$.
\end{theorem}

Theorem~\ref{thm:ce-complete} states that we can discover every potential bug
and construct a counterexample for it,
assuming the underlying solver is complete
for queries on first-order data.

\begin{theorem}[Relative Completeness of Counterexamples]
\label{thm:ce-complete}\ \\
If $\spair\mexpo{\mstoo_1} \multistdstep \spair{\swrong\mlab\mop}{\mstoo_2}$
and $\spair\mexpo{\mstoo_1} \refines_{\vec\mlab} \spair\mexp{\msto_1}$
and $\mlab \in \vec\mlab$,
then $\spair\mexpo{\msto_1} \multistdstep \spair{\swrong\mlab\mop}{\msto_2}$
such that there is an instantiation $\mstoo$ to $\msto_2$.
\end{theorem}
\section{Extensions}
\label{sec:extensions}

We discuss important extensions to our system
for a more practical programming language
with dynamic typing, data structures, contracts, and mutable states.
In addition, we address the issue with termination.
Our end goal is apply the method to realistic Racket~\cite{dvanhorn:plt-tr1} programs.

\subsection{Dynamic typing}

Dynamically typed languages defer safety checks to run-time to
avoid conservative rejection of correct programs.
Such languages have mechanisms for run-time inspection of data's type tag.
We model this feature by extending primitive predicates with
run-time type tests such as \texttt{integer?} or \texttt{procedure?},
which operate in the same manner as \texttt{zero?} in the typed language.
Changes to the semantics are straightforward:
we insert a run-time check into each application
to ensure a function is begin applied,
and into each primitive application to ensure arguments
have the right tags.
We also modify the rules for applying unknown functions,
where previous static distinction in function types
are turned to corresponding dynamic checks.

\subsection{User-defined data structures}
We extend the semantics to allow user-defined data structures,
enabling programmers to express rich data such as lists and trees.
Below is an example definition of a binary tree's node:
{\small \begin{alltt}
        (struct node (left content right))
\end{alltt}}
Each field in a data structure may itself be another data structure,
function, or base value.
Following the same treatment as functions,
we do not encode data structures in the solver.
Instead, we rely on execution to incrementally refine
an unknown value's shape when knowing that it has a specific tag.
For example, an unknown \texttt{node} has the shape of
\texttt{(node \(\slab\sb{1}\) \(\slab\sb{2}\) \(\slab\sb{3}\))}
where each of the fields $\slab_1$, $\slab_2$ and $\slab_3$
is an unknown and refinable value.
As before, we only need to encode constraints on
base values at the leaves of data structures.

\subsection{Contracts}
\label{sec:ext-contract}
Contracts generalize pre-and-post conditions to higher-order
specifications~\cite{dvanhorn:Findler2002Contracts},
allowing programmers to express rich invariants using arbitrary code.
They can either refine an existing type system~\cite{dvanhorn:Hinze2006Typed}
or ensure safety in an untyped language.

The following Racket~\cite{dvanhorn:plt-tr1} program illustrates
the use of a higher-order contract.
Function \texttt{argmin} requires a number-producing function
as its first argument and a list as its second,
and returns the list's element that minimizes the function.
{\footnotesize
\begin{alltt}
;(argmin f xs) \(\rightarrow\) any/c
;  f  : (any/c \(\rightarrow\) number?)
;  xs : (and/c pair? list?)
(define (argmin f xs)
  (argmin/acc f (car xs) (f (car xs)) (cdr xs)))

(define (argmin/acc f b a xs)
  (cond
   [(null? xs) a]
   [(< b (f (car xs))) (argmin/acc f a b (cdr xs))]
   [else (argmin/acc f (car xs) (f (car xs)) (cdr xs))]))
\end{alltt}}

Although the semantics of contract checking can be
complex~\cite{dvanhorn:Greenberg2010Contracts,dvanhorn:Dimoulas2011Correct},
it introduces no new challenges in our system.  We simply rely on the
semantics of contract checking itself to break down complex and
higher-order contracts into simple predicates.  In addition, opaque
flat contracts can be modeled soundly and precisely by rules for
opaque application.  Extension to the contract checking semantics
enables our system to construct counterexamples to violated
higher-order contracts.

\subsection{Mutable state}
We support stateful programs by extending the language with primitives
for assigning to and dereferencing mutable cells, along with a type
tag predicate.

When an unknown function is applied to a mutable cell,
it may invoke its content and mutate the cell arbitrarily.
Second, if its argument is a function, the unknown context may apply
the function any number of times, affecting the argument's internal state arbitrarily.
Finally, in the presence of mutable states, the system can no longer
assume that each function yields equal outputs for equal inputs,
so a memoized mapping is no longer applicable.
Because this last change can be too conservative for reasoning about idiomatic functional programs,
where programmers often think of functions as pure and use mutable cells judiciously,
it is useful in a practical system to have a special annotation for marking an unknown
function as pure.\footnote{First-class contracts can have internal states and enforce extensionality,
which symbolic execution can make use of.}

One challenge introduced by mutable cells is aliasing.
For example, a result from applying an unknown function can either
be a fresh value, or any previous value on the heap.
Future side effects performed on this unknown result may or may not affect
an existing mutable cell.
To soundly execute symbolic programs with mutable states,
we modify the behavior of primitive $\sboxp$ for run-time testing
of mutable cells.
When an unknown value $\mlab$ is determined to be a mutable cell,
it is non-deterministically a distinct cell from any previous one on the heap
or an alias to each previous cell and perform a substitution in the entire program.
Although this process is expensive, mutable cells are sparse in idiomatic Racket programs.
Programs with no invocation of $\sboxp$ (which is implicit in other operations)
do not pay this cost.
More efficient handling of aliasing in large imperative programs is one direction of our future work.

\subsection{Termination}

The semantics presented so far does not guarantee termination.
We can either accelerate (but not guarantee) termination
by detecting recursion and widen values accordingly~\cite{dvanhorn:Nguyen2014Soft},
or guarantee termination through systematic transformation of the semantics
into a finite state or pushdown analysis of itself~\cite{dvanhorn:VanHorn2010Abstracting}.
These techniques introduce spurious paths as over-approximations
to actual execution branches.
This affects both soundness and completeness of counterexamples.
First, it requires more work to guarantee soundness.
Because multiple concrete traces may be approximated to the same abstract trace,
running the program with one instantiation of a constraint set
may steer the program's flow to a different concrete trace that has the same abstraction.
To ensure an instantiation corresponds to a real counterexample,
it is necessary to first run the program with the concrete value set
before reporting it as a counterexample.
Second, relative completeness is also lost in practice.
Even though execution still reveals every possible error,
approximation results in a less precise constraint set for each trace,
and the system may repeatedly query the solver for the wrong model
before timing out.
For example, a simplistic solver trying to refute that ``\texttt{factorial(n + 4)} $\geq$ \texttt{10}''
with no constraint may keep producing non-negative values for \texttt{n}.

Nevertheless, for our specific need of counterexample generation
to refine an existing verification system
(discussed next in section~\ref{sec:eval}),
we perform no abstraction.
We rely on the previous system
to prove the lack of counterexamples for a large set of correct programs
\cite{dvanhorn:Nguyen2014Soft}
(therefore, many correct programs without counterexamples do terminate).
When used in combination with a verification system,
abstracting the state space for counterexample generation is of little value,
and makes it difficult to later concretize values to obtain
a counterexample.

\section{Implementation and evaluation}
\label{sec:eval}

To evaluate our approach, we integrate counterexample generation into
an existing contract verification system for programs written
in a subset of Racket~\cite{dvanhorn:plt-tr1}.
The system previously either successfully verified correct programs
or conservatively reported probable contract violations
and did not distinguish definite program errors from potentially false positives.
With the new enhancement, the tool identifies
a subset of reported errors as definite bugs with concrete counterexamples.
Below, we describe implementation extensions, discuss promising experiment results,
and address current difficulties.

\subsection{Implementation}

Our implementation and benchmarks can be found at
\begin{center}
\href{https://github.com/philnguyen/soft-contract/tree/pldi-2015}
{\tt github.com/philnguyen/soft-contract/tree/pldi-2015}
\end{center}
The prototype handles a much more realistic set of language features beyond SPCF.
First, our implementation supports dynamic typing with user-defined structures and first
class contracts as discussed in section~\ref{sec:extensions}.
We also support more contract combinators such as conjunction, disjunction,
and recursion.
Second, we extend the set of base values and primitive operations,
such as pairs, strings and Racket's full numeric tower,
which introduces more error sources and interesting counterexamples.
Finally, we employ a module system to let users organize code.
A module can export multiple values and define private ones for internal use.

Apart from being implemented as a command line tool,
our prototype is also available as a web REPL at
\begin{center}
\href{http://scv.umiacs.umd.edu}{\tt scv.umiacs.umd.edu}
\end{center}
The system attempts to verify correct programs and refute erroneous programs
with concrete counterexamples.
In some cases, it reports a probable contract violation without
giving any counterexample due to limitations of the underlying solver,
or the server simply times out after 10 seconds.

\subsection{Evaluation}

We collect benchmarks for our analysis from two sources: (1) prior
published work and (2) submissions to the web REPL we built.

Benchmarks from prior work are drawn from research on higher-order
model checking~\cite{dvanhorn:Kobayashi2011Predicate}, dependent type
checking~\cite{dvanhorn:Terauchi2010Dependent}, occurrence type
checking~\cite{dvanhorn:TobinHochstadt2010Logical}, and our own work
on contract verification~\cite{dvanhorn:Nguyen2014Soft}.  Since these
prior works focus on verification, the benchmarks are largely correct
programs.  In order to evaluate our counterexample generation
technique, we modify each of the programs to introduce errors.  To do
so, we weakened preconditions and omitted checks before performing
partial operations.  For example, a resulting program may deconstruct
a potentially empty list or compare potentially non-real numbers.  We
believe these changes are representative of common mistakes.  A
complete listing of the modifications is
available.\footnote{\url{https://github.com/philnguyen/soft-contract/blob/pldi-2015/soft-contract/benchmark-verification/diff.txt}}

Benchmarks from our web service are submitted (anonymously) by users
experimenting with the verification system.  Many of these programs
are buggy and we test how effective at discovering counterexamples.

In total, the evaluation is run on 282 programs consisting of 4050
lines of Racket code, excluding empty lines and comments.  The largest
programs are three student video games of the order of 250 lines.  The
test suite includes correct programs the system tries to verify as
well as incorrect programs the system tries to generate
counterexamples for.

We summarize our benchmark results in table~\ref{tbl:benchmarks}.
Each row shows the program's size (column 1),
its highest function order (column 2),
the time taken to verify the correct version of the program (column 3, if applicable),
and the time taken to generate a counterexample refuting an incorrect variant of the program
(column 4, if applicable).
We compute each program's order by inspecting its contract's syntax
(which is an under-approximation, because a contract may be dynamically computed).
The last 3 rows ``others'', ``others-e'' and ``others-w'' summarize many small programs
from our own benchmark suite as well as those collected from the server;
we report their total, minimum and maximum line counts,
total verification time, and highest function orders.
With the exception of 5 programs in the last row ``others-w'',
the system gives a counterexample for each incorrect program
in a reasonable amount of time:
the most complicated error takes 7 seconds to detect,
and most errors in typical higher-order programs take less than 2 seconds.
The last row shows benchmarks (all contributed by anonymous users)
that reveal the limitation of our counterexample generation in practice.
In each of these cases, the system soundly reports a probable contract violation,
but is unable to generate a counterexample confirming it.
We discuss current shortcomings and language features known
to thwart the tool in section~\ref{sec:difficulties}.

The overall result is promising.
First, there are specific examples where our prototype proves to be
a good complement to random testing.
For example, the tool finds a counterexample to the following program quickly and automatically:
\begin{alltt}
    f n = (/ 1 (- 100 n))
\end{alltt}
By default, QuickCheck does not find this error as it only considers
integers from \texttt{-99} to \texttt{99}.
Because QuickCheck treats a program as a black box,
this conservative choice is reasonable for fear that the integer may be
a loop variable causing the test case to run for a long time~\cite{local:Hughes2015QuickCheck}.
In contrast, our method explores the program's semantics
symbolically and discovers \texttt{100} as a good test case.

Second, the resulting higher-order counterexamples suggest that
the analysis can produce useful feedback.
For example, it is easy for programmers to forget that Racket supports
the full numeric tower~\cite{dvanhorn:StAmour2012Typing}
and that the predicate \texttt{number?} accepts complex numbers.
The contract on \texttt{argmin} in section~\ref{sec:ext-contract}
is in fact too weak to protect the function.
The system proves \texttt{argmin} unsafe by applying it to a specific
combination of arguments.
First, \texttt{f} is given a function that produces a non-real number,
which causes \texttt{<} to signal an error.
Second, \texttt{xs} is given a list of length 2, which is
the minimum length to trigger a use of \texttt{<}.
\begin{alltt}
    f = \syntax{(λ (x) 0+1i)}; xs = \syntax{(list 0 0)}
\end{alltt}

Finally, the tool analyzes the functional encoding of object-oriented programs effectively.
Zombie is one such example with extensive use of higher-order functions to encode objects and classes,
and the analysis can reveal errors buried in delayed function calls.
We believe this is a promising first step for generating classes and
objects as counterexamples.
In the example below, we define interface \texttt{posn/c}
that accepts two messages \texttt{x} and \texttt{y},
and function \texttt{first-quadrant?} that tests whether a position
is in the first quadrant.
\begin{alltt}
  (define posn/c
    ([msg : (one-of/c 'x 'y)]
     \(\rightarrow\) (match msg ['x number?] ['y number?])))

  ; posn/c \(\rightarrow\) boolean?
  (define (first-quadrant? p)
    (and (\(\ge\) (p 'x) 0) (\(\ge\) (p 'y) 0)))
\end{alltt}
The counterexample reveals one conforming implementation to interface
\texttt{posn/c} that causes error in the module.
\begin{alltt}
    (λ (msg) (case msg [(x) 0+1i] [(y) 0]))
\end{alltt}

\begin{table}[t]
\begin{center}
\begin{tabular}{|l|r|r|r|r|}
{\scriptsize Program} & {\scriptsize Lines} & {\scriptsize Order} & {\scriptsize Correct {\tiny (ms)}} & {\scriptsize Incorrect {\tiny (ms)}} \\
\hline
\multicolumn{5}{|c|}{\scriptsize Kobayashi et al. 2011 benchmarks}\\
  fhnhn & 18 & 2 & 38 & 50 \\
  fold-div & 18 & 2 & 321 & 160 \\
  fold-fun-list & 20 & 3 & 92 & 442 \\
  hors & 25 & 2 & 49 & 34 \\
  hrec & 9 & 2 & 52 & 143 \\
  intro1 & 13 & 2 & 24 & 128 \\
  intro2 & 13 & 2 & 25 & 127 \\
  intro3 & 13 & 2 & 25 & 23 \\
  isnil & 9 & 1 & 13 & 9 \\
  max & 14 & 2 & 32 & 135 \\
  mem & 12 & 1 & 22 & 254 \\
  mult & 9 & 1 & 61 & 147 \\
  nth0 & 15 & 1 & 19 & 296 \\
  r-file & 50 & 1 & 74 & 123 \\
  r-lock & 17 & 1 & 56 & 49 \\
  reverse & 11 & 1 & 15 & 205 \\
\hline
\multicolumn{5}{|c|}{\scriptsize Terauchi 2010 benchmarks}\\
  boolflip & 10 & 1 & 10 & 22 \\
  mult-all & 10 & 1 & 9 & 225 \\
  mult-cps & 12 & 1 & 253 & 35 \\
  mult & 10 & 1 & 72 & 21 \\
  sum-acm & 10 & 1 & 33 & 833 \\
  sum-all & 9 & 1 & 8 & 186 \\
  sum & 8 & 1 & 44 & 19 \\
\hline
\multicolumn{5}{|c|}{\scriptsize Tobin-Hochstadt and Felleisen 2010 benchmarks}\\
  {\small occurrence (14)} & 116 & 1 & 99 & 226 \\
\hline
\multicolumn{5}{|c|}{\scriptsize Nguy\^{e}n et al. 2014 benchmarks (video games)}\\
  snake & 164 & 1 & 37,350 & 2,476 \\
  tetris & 267 & 2 & 11,809 & 2,188 \\
  zombie & 249 & 3 & 19,239 & 954 \\
\hline
\multicolumn{5}{|c|}{\scriptsize Nguy\^{e}n et al. 2014 other benchmarks and anonymous web submissions}\\
  {\small others (73)} & {\scriptsize (2 - 51)} 818 & 3 & 20,465 & - \\
  {\small others-e (124)} & {\scriptsize (3 - 23)} 972 & 3 & - & 19,588 \\
  {\small others-w (5)} & {\scriptsize (4 - 4)} 20 & 1 & - & 431* \\
\hline
\end{tabular}
\end{center}
\caption{Program verification and refutation time}
\label{tbl:benchmarks}
\end{table}

\subsection{Difficulties}
\label{sec:difficulties}

We discuss current difficulties to our approach
and solutions in mitigating them.

First, the analysis is prone to combinatorial explosion
as inherent in symbolic execution.
Our tool finds bugs by performing a simple breadth-first search
on the execution graph, then stops and reports on the first
error encountered with a fully concrete counterexample.
In practice, most conditionals come from case analyses
instead of independent alternatives, and we rely on a precise
proof system to eliminate spurious paths.
In addition, the modularity mitigates the problem further,
as modules tend to be small, and contracts at boundaries
help recovering necessary precision.

One major source of slowdown in our system is complex preconditions,
where each input is guarded against a deep, inductively defined property.
Execution follows different branches before being able to generate
a valid input to continue verifying the module.
A naive breadth-first search is bogged down by a large frontier
resulting from different attempts to generate inputs,
most of which are eventually found invalid.
%
%
To mitigate this slow-down, we identify a class of expressions
as likely to lead to counterexamples and prioritize their execution.
Specifically, an expression whose innermost contract monitoring
is of a first-order property on a concrete module
is likely to reveal a bug.\footnote{In a symbolic program, the monitored value in this position
is usually abstract and covers all values the module produces.}
In contrast, expressions in the middle of input generation do not have this form,
because the inner-most contract monitoring is on the opaque input source.
Once the system successfully instantiates a concrete input and turns the
program into this ``suspect'' form,
it focuses on exploring this branch with that input instead of
trying numerous other inputs in parallel.
Using this simple heuristic, we are able to cut
the execution time of a module violating the ``braun-tree'' invariant
from non-terminating after 1 hour down to 2 seconds.

Second, there is a mismatch in the data-types between Z3's data-type and Racket's rich numeric tower.
In particular, Racket supports mixed arithmetic between different types of numbers
up to complex numbers~\cite{dvanhorn:StAmour2012Typing},
while Z3's treatment of numbers resembles that from most statically typed languages,
and the solver does not perform well in generating models involving a dynamic restriction
of a number's type.
Below is an example in the last row in table~\ref{tbl:benchmarks}
where the tool fails to generate a counterexample:
\begin{alltt}
    ; (integer? \(\rightarrow\) integer?)
    (define (f n) (/ 1 (+ 1 (* n n))))
\end{alltt}
In Racket, division is defined on the full numeric tower,
and the result of \texttt{(/ 1 (+ 1 (* n n)))} may not be an integer.
In the generated query, this result
is an unknown number $\slab$ of type \texttt{Real},
and the solver cannot give a model to a constraint set
asserting ``\texttt{(not (is\_int \(\slab\)))}''.
In addition, Racket distinguishes between exact and inexact numbers,
where inexact numbers are floating point approximations.
Because Z3 does not reason about floating points,
we currently do not soundly model inexact arithmetic.

\section{Related work}
\label{sec:related}

We relate our work to four main lines of research:
symbolic execution,
counterexample guided abstraction refinement for dependent type inference,
random testing, and contract verification.

\paragraph{First-order Symbolic Execution}
Symbolic execution on first-order programs is mature and has been used
to find bugs in real-world programs~\cite{local:Cadar2006EXE,local:klee08}.
\citet{local:Cadar2006EXE} presents a symbolic execution engine for C
that generates counterexamples of the form of mappings from addresses to bit-vectors.
Later work extends the technique to generate comprehensive test cases that discover bugs
in large programs interacting with the environment~\cite{local:klee08}.

\paragraph{Counterexample-guided Abstraction Refinement}
CEGAR has been used in
model checking and dependent type
inference~\cite{dvanhorn:Rondon2008Liquid,dvanhorn:Kobayashi2011Predicate,dvanhorn:DBLP:conf/vmcai/ZhuJ13},
where the inference algorithm iteratively uses a counterexample given by the solver
to refine preconditions attached to functions and values.
In case the algorithm fails to infer a specification,
the counterexample serves as a witness to a breaking input.
Our work finds higher-order counterexamples only by integrating a first-order solver,
and is applicable to both typed and untyped languages.
In contrast, dependent type inference relies on an extension to ML.
In addition, work on higher-order model checking
analyzes complete programs with first-order unknown inputs,
while we analyze partial programs with potentially higher-order unknown
values at the boundaries.

\paragraph{Random Testing}
Random testing is a lightweight technique for finding counterexamples to
program specifications through randomly generated inputs.
QuickCheck for Haskell~\cite{dvanhorn:Claessen2000QuickCheck} proves the approach
highly practical in finding bugs for functional programs.
Later works extend random testing to improve code coverage and scale
the technique to more language features such as states and class systems.
\citet{dvanhorn:Heidegger2010ContractDriven} use contracts to guide
random testing for Javascript, allowing users to annotate inputs to
combine different analyses for increasing the probability of hitting
branches with highly constrained preconditions.
\citet{dvanhorn:Klein2010Random} also extend random testing to work on
higher-order stateful programs, discovering many bugs in
object-oriented programs in Racket.
\citet{dvanhorn:Seidel2014Type} use refinement types as generators for
tests, significantly improving code coverage.

Our approach is a complement to random testing.
By combining symbolic execution with an SMT solver, the method takes advantage of
conditions generated by ordinary program code and not just user-annotated contracts.
In addition, the approach works well with highly constrained preconditions without further
help from users.
In contrast, random testing systems typically require programmers to
implement custom generators~\cite{dvanhorn:Claessen2000QuickCheck}
or require user annotations to incorporate a specific analysis collecting all literals
in the program to guide input construction~\cite{dvanhorn:Heidegger2010ContractDriven}.
Type-targeted testing~\cite{dvanhorn:Seidel2014Type}
is more lightweight and does not necessitate an extension to the existing semantics,
but gives no guarantee about completeness, as inherent in random testing.
Even though the tool rules out test cases that fail the pre-conditions,
regular code and post-conditions do not help the test generation process.
Our system makes use of both contracts and regular code to guide the execution
to seek inputs that both satisfy pre-conditions and fail post-conditions.
Exploring possible combination of symbolic execution and random testing
for more efficient bug-finding in higher-order programs is our future work.


\paragraph{Contract Verification and Refinement Type Checking}
Contracts and refinement types are mechanisms for specifying much richer
program invariants than those allowed in a typical type system.
Verification systems either restrict the language of refinements to be
decidable~\cite{dvanhorn:Rondon2008Liquid}
or allow arbitrary enforcement but leave unverifiable invariants as residual
run-time checks~\cite{dvanhorn:Flanagan2006Hybrid,dvanhorn:Knowles2010Hybrid,dvanhorn:Xu2012Hybrid,dvanhorn:TobinHochstadt2012Higherorder}.
While verification proves the absence of errors but may give false positives,
our tool aims to discover concrete, real counterexamples to faulty programs.
Our work is a direct extension to previous work on symbolic execution of
higher-order programs~\cite{dvanhorn:TobinHochstadt2012Higherorder}
and can be viewed as a complement to contract verification.

\section{Conclusion}
\label{sec:conclusion}

We have presented a symbolic execution semantics for finding concrete
counterexamples in higher-order programs and proved it to be sound and
relatively complete.  An early prototype shows that the approach can
scale to realistically sized functional programs with practical
features such as first-class contracts.  From the programmer's
perspective, the approach is lightweight and requires no custom
annotation to get started.  However, if contracts are present, they
can help guide the search for counterexamples.  Combined with previous
work on contract verification, it is possible to construct a tool that
can statically guarantee contract correctness of programs and
simultaneously ease the understanding of faulty programs, speeding up
the development of reliable software.

\ifpreprint \relax
\else

\acks We thank Sam Tobin-Hochstadt for countless discussions that
contributed significantly to the development of this work. Robby
Findler provided considerable feedback that improved the presentation
of this paper.  Clayton Mentzer helped build the web REPL and Andrew
Reuf gave valuable guidance on our artifact submission. We thank John
Hughes and Suresh Jagannathan for comments on prior work.  We thank
the PLDI and PLDI ERC reviewers for their detailed reviews, which
helped to improve the presentation and technical content of the paper
and accompanying artifact.  This research is supported in part by the
National Security Agency under the Science of Security program.  \fi

\balance
\bibliographystyle{abbrvnat}
\bibliography{dvh-bibliography,local}

\end{document}